\documentclass[12pt]{iopart}
\pdfoutput=1
\usepackage{textcomp}
\usepackage{iopams}
\usepackage{setstack}
\usepackage{amsthm}
\usepackage{graphicx}
\usepackage{epstopdf}
\usepackage{soul}
\usepackage{listings}
\usepackage[labelsep=period,font={footnotesize},labelfont={bf}]{caption}
\usepackage[font={scriptsize}, labelformat={simple}]{subcaption}
\usepackage{verbatim}
\usepackage{alltt}
\usepackage{mathrsfs}
\usepackage{color}
\usepackage[usenames,dvipsnames,svgnames,table]{xcolor}
\usepackage{rotating}
\usepackage{chngpage}

\usepackage[numbers,square,comma,sort&compress]{natbib}
\usepackage[plainpages=false,pdfpagelabels]{hyperref}

\hypersetup{%
  pdftitle={W44+ EIE},
    pdfauthor={M M Bluteau},
    pdfkeywords={},
    colorlinks = true,
    linkcolor = black,
    anchorcolor = black,
    citecolor = black,
    filecolor = black,
    urlcolor = black,
    pdfprintscaling=None
}

\newcommand{\rmatrix}{\ensuremath{R}-matrix}
\newcommand{\grasp}{{\sc grasp\ensuremath{^0}}}
\newcommand{\AS}{{\sc autostructure}}
\newcommand{\darc}{{\sc darc}}
\newcommand{\Wff}{W\ensuremath{^{44+}}}

\newcommand{\PEC}{\ensuremath{\mathcal{PEC}}}
\newcommand{\transarraya}{[3d$^{10}$4s$^{2}$--3d$^9$4s$^2$4f]}
\newcommand{\transarrayb}{[3d$^{10}$4s$^{2}$--3d$^9$4s4p4d]}
\newcommand{\helvet}[1]{{\fontfamily{phv}\selectfont #1}}
\newcommand{\dd}{\ensuremath{\mathrm{d}}}
\newcommand{\Te}{\ensuremath{T_{\mathrm{e}}}}
\renewcommand{\PL}{\ensuremath{P_{L,1, j \rightarrow k}}}
\newcommand{\PLT}{\ensuremath{P_{LT,1}}}
\newcommand{\FPEC}{\ensuremath{\mathcal{F}\textrm{-}\mathcal{PEC}}}
\newcommand{\nfive}{3d$^{10}$4$l$5$l^{\prime}$}

\begin{document}
\newif\ifrough
\roughfalse

\title[\Wff EIE]{Dirac \rmatrix \ and Breit-Pauli 
  distorted wave calculations of the electron-impact excitation of \Wff}
\author{M M Bluteau, M G O'Mullane, and N R Badnell}
\address{Department of Physics, University of Strathclyde, Glasgow G4 0NG, UK}
\ead{matthew.bluteau@strath.ac.uk}

\begin{abstract}
  With construction of ITER progressing and existing tokamaks carrying-out ITER-relevant
  experiments, accurate fundamental and derived atomic data for numerous
  ionization stages of tungsten (W) is required to assess the potential effect
  of this species upon fusion plasmas. The results of fully relativistic,
  partially radiation damped, Dirac \rmatrix\ electron-impact excitation calculations for
  the \Wff{} ion are presented. These calculations use a configuration
  interaction and close-coupling expansion which opens-up the
  3d-subshell, which does not appear to have been considered before
  in a collision calculation. As a result, it is possible to
  investigate the arrays, \transarraya\ and \transarrayb, which are predicted
  to contain transitions of diagnostic importance for the soft x-ray region.
  Our \rmatrix{} collision data are compared with previous \rmatrix{} results
  by Ballance and Griffin as well as our own relativistically corrected,
  Breit-Pauli distorted wave and plane-wave Born calculations. All relevant
  data are applied to the collisional-radiative modelling of atomic populations,
  for further comparison. This reveals the paramount nature of the 3d-subshell
  transitions from the perspectives of radiated power loss and detailed
  spectroscopy.  
  
\end{abstract}


\section{Introduction} 
\label{sec:Introduction}

One of the obstacles that ITER and future magnetic confinement fusion (MCF)
devices must overcome is the resilience and impact of erosion of the plasma
facing components (PFCs). Tungsten (W) metal is currently a top candidate owing
to its advantageous thermo-mechanical properties: a high melting point and
heat-load capacity, a low sputtering rate \cite{bolt2004}, and a low rate of
tritium co-deposition compared to impurities from carbon based PFCs
\cite{federici2001}. ITER will now only use a full-W divertor \cite{pitts2013,
aymar2001, aymar2002}. As a result, elemental W will inevitably enter the
fusion plasma by physical sputtering or evaporation \cite{iter1999a}, and the
consequences of this can be mixed.  With its large atomic number, $Z=74$, W has
the potential to achieve high residual charge states, $z = Z - N$, where $N$ is 
the number of electrons.  Because of the $(z+1)^4$ scaling of dipole, $\Delta n > 0$
radiative rates, W ions have an increased propensity to undergo radiative
transitions compared to low-$Z$ species in the same isoelectronic sequence. In
other words, impurity W ions are efficient at radiating their energy and can
greatly contribute to radiative power loss from the plasma: emission line power
losses, which have a $(z+1)$ scaling dependent on the type and relative energy of
the transition, will dominate over bremsstrahlung in this context. Significant
modelling from an atomic physics perspective will be necessary to quantify the
impact of radiation losses due to W ions.

Fortunately, it is not necessary to consider all ionization stages of W in
depth. Even modern devices have insufficient temperatures to fully ionize W,
and only certain ions will be present at different locations in the plasma
vessel. The important ionization stages will be determined by operating
parameters of the device. \Wff\ is an ion of interest for spectral diagnostics
on JET and is located in the core of the tokamak plasma. Spectral lines in the
soft x-ray region have been observed by the bent crystal x-ray spectrometer,
KX1 \cite{shumack2014}. For \Wff, lines in this region are produced by
transitions to the 3d-subshell. In particular, the transitions in the
\transarraya{} and \transarrayb{} arrays are dominant because the upper levels
are populated directly by excitation from the ground, as summarized in
\cite{kramida2009}.  Lines for these transitions have been observed
experimentally using electron-beam ion traps (EBITs) 
\cite{kramida2011, kramida2009, clementson2010,
neill2004}, and theoretical atomic structure calculations by Fournier
\cite{fournier1998} and Spencer et al \cite{spencer2014} confirm large oscillator strengths. However, to our
knowledge, no collision calculation or spectral modelling gives a complete
consideration to both of these obviously important, 3d-subshell transition
arrays, so the primary objective of the present work is to rectify this shortcoming. 
Two wavelengths in particular will seen to be relevant: 5.76 \AA\ and 5.94 \AA.  

A fairly recent work on the ionization balance of the W isonuclear sequence was
conducted by P\"utterich et al.~\cite{puetterich2008}; therefore, we focus on
considerations for another important spectral modelling quantity, the Photon
Emissivity Coefficient ($\mathcal{PEC}$, explained in section \ref{sub:Atomic
Population Modelling}).  To obtain the relevant \PEC\ data, it is necessary to
generate fundamental atomic data for the various processes connecting the
levels of the ion or atom.  In fusion plasmas, the dominating excitation
process is electron-impact excitation (EIE). 
To improve upon our current plane-wave Born (PWB) baseline
calculations, a full close-coupling (CC) approach should be used, and due to
the high residual charge of \Wff, $z = 44$, the effect of radiation damping
of resonances should also be considered \cite{robicheaux1995}.  Moreover, relativistic
effects must be incorporated by one means or another due to the high nuclear
charge, and the 3d-subshell transitions motivated above must be included.
Prior to the present collision calculations, no data in the literature satisfied all of
these conditions; however, there have been limited EIE calculations for \Wff\ with
which we will benchmark.

Previous relativistic \rmatrix\ calculations have been conducted by Ballance
and Griffin \cite{ballance2007} using essentially the same codes employed in
this study, and it is with their results that we seek to compare. However,
their calculations do not include any configurations involving excitation from
the 3d-subshell, which constitutes a serious shortcoming from our present
perspective and is the primary motivation for this study. (It should be
noted that the importance of opening-up the 3d-subshell for diagnostic
purposes was not appreciated until we carried-out a preliminary survey
of what might constitute the main emission lines.)
Conversely, the Ballance and Griffin calculations do include a full treatment of all 
types of radiation damping, whereas the current study only contains a partial treatment. 
Our reasons for including only the core radiation of Rydberg resonances (type-I damping)
 are detailed in Section
\ref{sub:DARC Execution}. Additionally, Das et al.~have conducted fully
relativistic distorted wave (DW) calculations for \Wff and other W ions in
\cite{das2012}. This study does not satisfy our criterion of using a CC, and
more importantly, it omits a crucial configuration, 3d$^9$4s4p4d, 
the effect of which is further investigated in section 
\ref{sub:CI and Structure Determination}.  

We seek to fill the gap in \Wff\ EIE data with fully relativistic,
partially damped, Dirac \rmatrix\ calculations conducted using \darc{} (see
section \ref{sub:DARC Execution}). These calculations include configurations
with a 3d-hole so that the \transarraya\ and \transarrayb\ transition arrays
are accommodated. \AS\ was also employed in various capacities to
support these calculations, including its Breit-Pauli distorted wave (BPDW)
approach for generating EIE data. Ultimately, a proper 
spectral modelling of the \Wff{} spectrum with particular attention to the
3d-subshell transitions for verification of
their importance is needed. This modelling will be conducted through use of the
Atomic Data and Analysis Structure (ADAS) \cite{adas}, facilitating future
comparison with experiment. 

The structure of the remainder of the paper is as follows. Section
\ref{sec:Methodology} describes the methodology used to conduct the
calculations, and it is divided into four subsections. First, section
\ref{sub:CI and Structure Determination} lists and explains the specification
of the configuration interaction (CI), which is critical for an accurate
investigation of the 3d-subshell transitions and differentiates the present results
from previous works. Second, section \ref{sub:DARC Execution} provides
the necessary technical and physics details for our use of the \darc{} and
\AS\ codes. Third, section \ref{sub:AUTOSTRUCTURE DW and Born Limits}
discusses some important issues regarding infinite energy collision strength limits
in \darc{}. Lastly, section \ref{sub:Atomic Population Modelling} provides some
background and technical details for the atomic population modelling carried
out in this study. Section \ref{sec:Results and Discussion} presents the
results of the present calculations along with the relevant analysis in three sections:
atomic structure, collision data, and atomic population modelling. Finally, the present
work is summarized and future options considered in section
\ref{sec:Conclusion}.



\section{Methodology} 
\label{sec:Methodology}

\subsection{CI and Structure Determination} 
\label{sub:CI and Structure Determination}

Our focussed consideration of the 3d-subshell transition arrays, \transarraya{}
and \transarrayb{}, requires the inclusion of configurations with a 3d-hole.
Apart from the 3d$^9$4s$^2$4f and 3d$^9$4s4p4d configurations, there are
several other configurations to consider due to the possibility of mixing, and
it was not immediately obvious which ones should have been included in the
configuration interaction (CI) of the target structure calculation.  One must
be prudent in selecting the CI due to computer memory limits at the collision
calculation stage: a compromise between the number of $J$-resolved
levels and the overall accuracy of results must be reached. Two structure codes
were employed at this junction: \AS{}\footnote{Version 24.24} \cite{badnell1986,badnell1997,badnell2011b},
which uses the Breit-Pauli Hamiltonian and nonrelativistic wavefunctions and
\grasp\ \cite{grant1980,McKenzie1980,dyall1989,graspdarc2009}, which uses the
Dirac Hamiltonian (with the Breit interaction) and Dirac-Fock spinors. The
final CI included 13 configurations and resulted in 313 $LSJ\pi$ levels, all
below the ionization limit: 
\begin{eqnarray*}
    3\textrm{d}^{10}4\textrm{s}^{2}, \ \ 3\textrm{d}^{10}4\textrm{s}4\textrm{p},
    \ \ 3\textrm{d}^{10}4\textrm{s}4\textrm{d},
    \ \ 3\textrm{d}^{10}4\textrm{s}4\textrm{f},\ \  3\textrm{d}^{10}4\textrm{p}^2,
    \ \ 3\textrm{d}^{10}4\textrm{p}4\textrm{d},\\
    3\textrm{d}^{10}4\textrm{p}4\textrm{f},  
    \ \ 3\textrm{d}^{10}4\textrm{d}^2,\ \  3\textrm{d}^{10}4\textrm{d}4\textrm{f},
    \ \ 3\textrm{d}^{9}4\textrm{s}^24\textrm{p},
    \ \ 3\textrm{d}^{9}4\textrm{s}^24\textrm{d},
    \ \ 3\textrm{d}^{9}4\textrm{s}^24\textrm{f},\\
    3\textrm{d}^94\textrm{s}4\textrm{p}4\textrm{d}.
    \label{eq:fullCI}
\end{eqnarray*}

Emphasis must be placed upon the 3d$^9$4s4p4d configuration, which has not been
considered in either structure or collision calculations until now, to the best
of our knowledge. As alluded to more generally in section
\ref{sec:Introduction}, it is because of this omitted configuration that a
proper modelling of the important 3d-subshell transition arrays for \Wff{} has
not been possible. The 3d$^9$4s4p4d configuration mixes heavily with
3d$^{9}$4s$^2$4f, and the subsequent effect upon the radiative data of the
dominant 3d-subshell transitions is presented in table \ref{tab:config_var}.
Observation of the changes between row 1 and row 2 clearly shows this effect,
and notably the ground to $^3$P$_1^{\circ}$ transition increases by 3 orders of
magnitude. Thus, comparison of the dominant 3d-subshell transitions between
calculations is only sensible if these calculations both include the
3d$^{9}$4s$^2$4f and 3d$^9$4s4p4d configurations. No further mention will be
made of the Das et al.~calculations for exactly this reason; they do not
include the 3d$^9$4s4p4d configuration, and preliminary comparison of our
collision data with theirs immediately revealed large discrepancies. It should
be noted that the effect of strong mixing between adjacent configurations
related by a promotion and demotion of $l$ quantum numbers has been well
documented in previous cases, such as Sn$^{10+}$ and Pr$^{21+}$
\cite{badnell2011a,bauche1987}. Table \ref{tab:config_var} also shows some
other candidate configurations that were omitted due to their lack of influence on
the radiative data: 3d$^{10}$4f$^2$, 3d$^9$4p$^3$, and 3d$^9$4s4p$^2$.

The primary calculation with which we compare is Ballance and Griffin's
\cite{ballance2007}, so it is important to rationalize the differences in the
CI basis sets. Row 4 contains the results for the union of the CI basis sets used in our calculations,
and it can be observed that the addition of the 3d$^{10}$4$l$5$l^{\prime}$
configurations do have a moderate effect on the 3d-subshell transitions
relative to row 2.  Ideally, all of these configurations should be included in
the CI and CC expansion, but the 397 levels generated by these configurations
is computationally inhibitive to the subsequent collision calculation. 
Because the soft x-ray, 3d-subshell transitions are
the focus of this study, the 3d$^{10}$4$l$5$l^{\prime}$ configurations had to
be omitted from our CI.  However, further influence of these configurations will
be assessed in section \ref{sub:Modelling_Results} by merging Ballance and Griffin's
\cite{ballance2007} data for the $n=5$ levels into our own dataset and observing the
effect upon the modelled results.

Our \grasp{} results closely mimic the \AS{} results in table
\ref{tab:config_var}. An extended average level (EAL) calculation, which
optimizes a weighted trace of the Hamiltonian matrix, was used for the \grasp\
calculation.  The target orbitals produced were used in the subsequent \darc{}
collision calculation, which is described in the section \ref{sub:DARC
Execution}. In addition, comparisons are made in section \ref{sub:Atomic
Population Modelling} to modelled results derived from our plane-wave
Born (PWB) calculations using Cowan's codes \cite{cowan1981}. The CI for these
calculations is slightly different, combining configurations from ours and
Ballance and Griffin's:
\begin{eqnarray*}
    3\textrm{d}^{10}4\textrm{s}^{2}, \ \
    3\textrm{d}^{10}4\textrm{s}4\textrm{p}, \ \
    3\textrm{d}^{10}4\textrm{s}4\textrm{d}, \ \
    3\textrm{d}^{10}4\textrm{s}4\textrm{f},\ \
    3\textrm{d}^{10}4\textrm{s}5\textrm{s}, \ \
    3\textrm{d}^{10}4\textrm{s}5\textrm{p},\\
    3\textrm{d}^{10}4\textrm{s}5\textrm{d},  \ \
    3\textrm{d}^{10}4\textrm{s}6\textrm{s},\ \
    3\textrm{d}^{10}4\textrm{s}6\textrm{p}, \ \
    3\textrm{d}^{10}4\textrm{s}6\textrm{d}, \ \
    3\textrm{d}^{9}4\textrm{s}^24\textrm{d}, \ \
    3\textrm{d}^{9}4\textrm{s}^24\textrm{f},\\
    3\textrm{p}^{5}3\textrm{d}^{10}4\textrm{s}^24\textrm{p}, \ \
    3\textrm{p}^{5}3\textrm{d}^{10}4\textrm{s}^24\textrm{d}.
    \label{eq:CowanCI}
\end{eqnarray*}
The aim of this CI basis set was to achieve more breadth of excited-state coverage.

\begin{table}[htbp]
    \caption{Summary of radiative data from \AS{} while varying the CI basis set. $A_{ki}$ is the
  Einstein A-coefficient (transition probability); $S_{ik}$ is the line
  strength; and $g_i f_{ik}$ is the weighted oscillator strength. The base 13
  configurations are those listed in section \ref{sub:CI and Structure
  Determination} but with
  3d$^9$4s4p4d replaced by 3d$^{10}$4f$^{2}$. All subsequent entries are for the 
  configurations that have been added or removed from this basis. BG07 refers
  to the configurations used in Ballance and Griffin's \Wff{} calculations
  \cite{ballance2007}.}
  \begingroup
  \fontsize{9pt}{9pt}\selectfont
\begin{center}
\begin{tabular}{llllllcllll}
\hline \hline
  \multicolumn{1}{c}{CI} & \multicolumn{1}{c}{\emph{k}} &
  \multicolumn{1}{c}{$i$} & \multicolumn{1}{c}{$A_{ki}$
  (s$^{-1}$)} &
  \multicolumn{1}{c}{$S_{ik}$(au)} &  
  \multicolumn{1}{c}{$g_i f_{ik}$} & \multicolumn{1}{c}{$(-1)^{\pi}(2S_k+1)$} & 
  \multicolumn{1}{c}{$L_k$} & \multicolumn{1}{c}{$J_k$} & 
  \multicolumn{1}{l}{$k$ conf.} & \multicolumn{1}{l}{Lvs} \\ \hline
base 13 & 126 & 1 & 1.31E+14 & 0.040392 &  2.07237 & -3 & 2 & 1 & 3d$^9$4s$^2$4f & 134 \\ 
        & 134 & 1 &  4.25E+14 &  0.118734 &   6.287 & -1 & 1 & 1 & 3d$^9$4s$^2$4f& \\ 
        & 116 & 1 & 1.16E+11 & 0.000038 &  0.00189 & -3 & 1 & 1 & 3d$^9$4s$^2$4f & \\ 
 \hline 
+3d$^9$4s4p4d
 & 288 & 1 &  1.11E+14 &  0.030694 &   1.63157 & -3 & 2 & 1 & 3d$^9$4s4p4d& 326 \\ 
 & 304 & 1 &  1.42E+14 &  0.038086 &   2.04257 & -1 & 1 & 1 & 3d$^9$4s$^2$4f& \\ 
 & 308 & 1 &  1.13E+14 &  0.030244 &   1.62454 & -3 & 1 & 1 & 3d$^9$4s4p4d & \\ 
 \hline 
+3d$^9$4s4p4d
 & 275 & 1 &  1.09E+14 &  0.030123 &   1.60008 & -3 & 2 & 1 & 3d$^9$4s4p4d& 313 \\ 
 --3d$^{10}$4f$^2$
 & 291 & 1 &  1.38E+14 &  0.037198 &   1.99348 & -1 & 1 & 1 & 3d$^9$4s$^2$4f& \\ 
 & 295 & 1 &  1.18E+14 &  0.031645 &   1.69855 & -3 & 1 & 1 & 3d$^9$4s4p4d & \\ 
 \hline
+3d$^9$4s4p4d
 & 359 & 1 &  1.05E+14 &  0.030072 &   1.58174 & -3 & 2 & 1 & 3d$^9$4s4p4d& 397 \\ 
 +BG07 (4$l$5$l^{\prime}$)
 & 374 & 1 &  1.11E+14 &  0.030929 &   1.64053 & -1 & 1 & 1 & 3d$^9$4s$^2$4f& \\ 
 & 388 & 1 &  1.15E+13 &  0.003136 &   0.16751 & -3 & 1 & 1 & 3d$^9$4s4p4d & \\ 
 \hline
+3d$^9$4s4p$^2$
 & 182 & 1 & 1.31E+14 & 0.040416 &  2.06936 & -3 & 2 & 1 & 3d$^9$4s$^2$4f & 190 \\ 
 & 190 & 1 &  4.21E+14 &  0.117906 &   6.24096 & -1 & 1 & 1 & 3d$^9$4s$^2$4f & \\ 
 & 172 & 1 & 1.16E+11 & 0.000037 &  0.00189 & -3 & 1 & 1 & 3d$^9$4s$^2$4f & \\ 
 \hline
+3d$^9$4p$^3$
 & 151 & 1 & 1.31E+14 & 0.04040 &  2.06867 & -3 & 2 & 1 & 3d$^9$4s$^2$4f & 172\\ 
 & 168 & 1 &  4.22E+14 &  0.11799 &   6.24569 & -1 & 1 & 1 & 3d$^9$4s$^2$4f & \\ 
 & 136 & 1 & 1.16E+11 & 0.000037 &  0.00189 & -3 & 1 & 1 & 3d$^9$4s$^2$4f & \\ 
 \hline
\end{tabular}
\end{center}
\endgroup
\label{tab:config_var}
\end{table}


\subsection{DARC and AUTOSTRUCTURE Execution} 
\label{sub:DARC Execution}

The Dirac \rmatrix, partially damped EIE results presented in this study were
generated using the \darc{} suite, developed by Norrington \cite{graspdarc2009}
and modified to incorporate parts of the parallel \rmatrix\ codes
\cite{mitnik2001, mitnik2003, ballance2004}.  Our calculational procedure is almost identical
to that described in \cite{ballance2006}; however, we did not perform a fully
damped calculation, as mentioned earlier, so the outer region calculation was
slightly different. 

If all possible types of radiation damping are to be accounted for, the bound
($N+1$)-electron eigenvalues, eigenvectors, and dipole matrix elements need to
be handled, which is a computationally expensive task. Moreover, because we
include configurations with an open 3d-subshell in our CI and CC expansion, the
number of levels in our calculation is nearly doubled compared to Ballance and
Griffin: 168 levels in their calculation versus 313 in the present one. As a consequence,
the computational demand of the present problem is greater initially, and it
is not practical to further expand the calculations by including all forms of
radiation damping at this point in time. However, the {\sc pstgf} outer
region code independently has the capability to include type-I damping via
Multichannel Quantum Defect Theory (MQDT) \cite{seaton1983} at minimal
computational cost. Type I damping constitutes the radiative transition of a
core, non-Rydberg electron starting from an intermediate, $(N+1)$-electron
resonance; type-I damping tends to dominate because of the $1/n^3$ scaling of
autoionization and Rydberg radiation rates. This is supported by our
results given in section \ref{sub:Collision Data}, and so our limited damping
approach is a suitable approximation. The outer region calculations were
run both with and without type-I damping.

The relevant physics parameters for the problem are as follows. The CI and
close-coupling (CC) expansion both incorporate all configurations determined in
section \ref{sub:CI and Structure Determination} resulting in
313 $LSJ\pi$ levels. Moreover,
although the calculations are already split into exchange and nonexchange
components at the spatial \rmatrix{} box boundary, they can be further
partitioned in angular momentum space, since exchange effects reduce at high
angular momentum values. Thus, a large $J$ value for the symmetries is selected
above which electron exchange effects can be neglected even in the inner
region; in the present case, full close-coupling equations were solved for $0.5\leq
J\leq16.5$ and the nonexchange versions for $17.5 \leq J \leq 35.5$. The actual
\rmatrix\ boundary is selected automatically such that all the bound orbitals
have magnitudes below an arbitrary threshold of $10^{-3}$; these settings
resulted in an \rmatrix\ boundary of 1.33 au. When specifying the generation
of continuum-electron orbitals, one should ensure that the energy range of
these orbitals for each angular momentum exceeds the intended range of
scattering electron energies by approximately a factor of 1.8 in practice. A
maximum scattering energy of 1100 Ryd was used for these calculations to match
Ballance and Griffin, and so the maximum energy eigenvalue of the
continuum-electron basis orbitals for a given angular momentum value should
exceed $\approx 1800$ Ryd. For the exchange case, this required 34 basis
orbitals per angular momentum value, and for the non-exchange case this required
30 basis orbitals per angular momentum.

The features of EIE collision strengths are dominated by intermediate
resonances in the energy range defined by transitions between target levels.
These resonances manifest as sharp and narrow peaks, meaning the collision
strengths need to be evaluated on a fine energy mesh in this region. The mesh
parameters used for the outer region code are summarized in table
\ref{tab:pstgf_mesh}. One will also note from table \ref{tab:pstgf_mesh} that a
further division has been introduced within the exchange case. Only for $J\Pi$
symmetries with $J\leq 8.5$ was the full fine mesh employed in the resonance
region. $\textrm{MXE} = 48000$ was chosen for this fine mesh in order to
closely mimic the number of points used in the previous \darc{} calculations by
Ballance and Griffin \cite{ballance2007}.

\begin{table}[htbp]
\caption{Summary of mesh cases and parameters for {\sc PSTGF}. MXE is the number
of points for the outer region energy mesh, and EINCR in the step size of the mesh
in Ryd/$z^2$. The resonance region is enclosed by the range, $[E_2 -
    E_{\textrm{incr}}, E_{313} + E_{\textrm{incr}}]$ and the high energy region
by $(E_{313} + E_{\textrm{incr}}, 1100 \textrm{ Ryd}]$. $E_i$ is the energy
eigenvalue of the $i$th excited level relative to the ground in Rydbergs:
$E_2 =6.34789294$ Ryd and $E_{313} =1.61979116\times 10^2$ Ryd}
\fontsize{8pt}{10pt}\selectfont
\begin{center}
\begin{tabular}{|l|l|l|l|}
\hline
\multicolumn{ 2}{|l|}{\textbf{Case}} & \textbf{Resonance Region} & 
\textbf{High Energy Region} \\ \hline
\multicolumn{ 1}{|l|}{Exchange} & $0 \leq J \leq 8.5$ & MXE=48000 EINCR=6.701E-06 & 
\multicolumn{ 1}{l|}{MXE=720 EINCR=0.0002562} \\ \cline{ 2- 3}
\multicolumn{ 1}{|l|}{} & $9.5 \leq J \leq 16.5$ & MXE=360 EINCR=0.0002252 & 
\multicolumn{ 1}{l|}{} \\ \hline
Nonexchange & $17.5 \leq J \leq 35.5$ & \multicolumn{ 2}{l|}{MXE=1008 EINCR=0.0002636} \\ \hline
\end{tabular}
\end{center}
\label{tab:pstgf_mesh}
\end{table}

In the interest of having more collision data for comparison, \AS{}
runs were also conducted using the same CI as for \darc{}/\grasp{}. The isolated
target structure calculation used an intermediate-coupling (IC) scheme with
relativistic, $\kappa$-averaged orbitals. Multi-electron interactions are
included through the Thomas-Fermi-Dirac-Amaldi model potential with scaling
orbital parameters, $\lambda_{nl}$, determined through a variational method of
all possible orbitals: 1s, 2s, 2p, 3s, 3p, 3d, 4s, 4p, 4d, 4f. The scattering
problem is solved using a Breit-Pauli distorted wave (BPDW) approach as
described in \cite{badnell2011b}. 

\subsection{Born Limits} 
\label{sub:AUTOSTRUCTURE DW and Born Limits}

It is important to give attention to the infinite energy limits of
collision strengths since their values correlate strongly with those of the
(background) collision strengths over a wide range of energies.
A limitation of the \darc{}/\grasp\ suite is that these infinite
energy limits are only calculated for the electric dipole-allowed transitions:
$\Delta J = \pm 1$ and parity change. 

To rectify this absence of data, the remaining calculated collision strength values are
extrapolated when convoluting. Because we cannot differentiate between transitions with Born
limits and those truly forbidden by selection rules, it is assumed the highest energy
calculated collision strength, $\Omega_f$, has nearly reached the infinite energy
limit, and so $\Omega_f$ is extrapolated as a constant. Although this is usually a good approximation, it relies
on calculating the collision strengths to an arbitrarily high energy.
Alternatively, the Born limits may be obtained from a different program and
spliced into the collision strengths file; a linear interpolation involving
this point can then be used.  However, because two different structure
calculations are being effectively combined, one must question how close the
structure calculations are and whether it even makes sense to combine the
results from different theories.

In the present case, the possibility of using the Born limits from our \AS{}
calculation was explored since Ballance and Griffin used Born limits from \AS{}
for their calculations \cite{ballance2007}. The only potential metric for
determining the suitability of the \AS\ Born limits is a comparison of the
(electric) dipole-allowed transition limits from \grasp{} and \AS{}. In practice, this is
simply a comparison of the line strengths --- see Burgess \& Tully
\cite{burgess1992}. A linear
comparison of the line strengths from the two codes reveals that only 24\% of
the transitions lie within 20\% of each other, with a mean percent difference
of 6185\% and a weighted mean percent difference of 11\%.  The weighting
factors, $w_{ik}$, are defined as
\begin{equation}
    w_{ik} = \frac{r_{ik}}{\sum_{j,l}r_{jl}}
    \ ;\quad r_{jl} = \log(S_{jl} \slash S_{\textrm{max}}) \ .
    \label{eq:weights}
\end{equation}
Based on this weighting scheme, the large discrepancy between the weighted and
unweighted means suggests that the differences between line strength values
tends to be relatively larger at lower magnitude line strengths. Indeed, this
supposition is supported by the observation of a linear scatter plot of the
line strengths, and it is a trend one might expect to see. Thus, the amount of
agreement between the \darc\ and \AS\ dipole limits depends on how
much importance one places upon the low and high magnitude values separately.

There is no reason to doubt that this behaviour would not also
extend to the Born limits; however, the effect would likely be exacerbated
since the average magnitudes of the infinite energy limits decreases by
approximately an order of magnitude for each subsequent multipole order. In the
absence of any Born limits from \grasp\ with which to compare, this less than
conclusive evidence from the dipole limits comparison does not resolve the
issue of whether any accuracy might be gained from splicing the \AS{}
Born limits. Given this uncertainty, we do not believe the effort of manually
tampering with the collision strength files is worthwhile, and so we retain
the default behaviour of extrapolating the high energy collision strengths
as constants for transitions without $E1$ dipole limits.

\subsection{Atomic Population Modelling} 
\label{sub:Atomic Population Modelling}
The total emissivity in a spectrum line (transition), $i \rightarrow k$, is
given by
\begin{equation}
    \varepsilon_{i \rightarrow k} = N_i A_{i\rightarrow k},
    \label{eq:emissivity}
\end{equation}
where $N_i$ is the population density of the upper state, $i$, in ionization
stage $z$ and $A_{i\rightarrow k}$ is the radiative transition rate from $i$ to
the lower state, $k$. The $A_{i\rightarrow k}$ values are straightforward to
obtain from the structure calculation for an ion; however, the $N_i$ require
some form of atomic population modelling. Just as for the fundamental EIE
cross-section data, full atomic population modelling that incorporates these
transitions is limited in the literature for \Wff.  Clementson et
al.~\cite{clementson2010} present the calculated spectrum for \Wff\ in an EBIT
plasma environment using a collisional-radiative model based on fundamental
data from FAC. Since these results are not applicable in the laboratory fusion
plasma regime, we plan to address this deficit in the \Wff\ modelled spectrum
data equipped with the new fundamental atomic data that incorporates the
dominant 3d-subshell transitions.

Our modelling of the $N_i$ employs \emph{collisional-radiative} (CR) theory and
the assumption that the lifetime of the ground state is far greater than any of
the excited states' lifetimes. This was determined based on preliminary
modelling that revealed collisional excitation from the metastable levels of
\Wff\ does not have a significant effect on excited state populations until an
electron density of $N_{\textrm{e}}\approx 10^{16}$ cm$^{-3}$, far outside the
parameter space of both current fusion devices and the proposed ITER limits
\cite{iter1999a}. It is the large energy separation amongst the metastables and
ground, caused by the large residual charge, $z=44$, that is responsible for
the absence of density effects in the current context. As a result, \emph{all}
atomic levels will be in \emph{quasi-static equilibrium} relative to the ground
state, which dominates the description of the species population. 

The population density of the ground is denoted by $N_1$, and the rate of
population density change of an excited state, $j$, is 
\begin{equation}
    \frac{\dd N_j}{\dd t} = C_{j1} N_{1} +
    \sum_{i} C_{j i} N_{i} \ .
    \label{eq:Npop}
\end{equation}
The $C_{j i}$ are elements of the collisional-radiative matrix and are
defined by 
\begin{equation}
    C_{j i} = A_{i\rightarrow j} \slash N_{\mathrm{e}} +
    q_{i\rightarrow j}^{\mathrm{e}},
    \label{eq:cr_elements}
\end{equation}
where $q_{i\rightarrow j}^{\mathrm{e}}$ is the electron-impact excitation or
de-excitation rate coefficient depending on the energy ordering of $i$ and
$j$. Enforcing
the quasi-equilibrium condition on the excited states ($\dd N_j\slash \dd t =
0$) and isolating for $N_i$ in \eref{eq:Npop}, one obtains
\begin{equation}
    N_i = -\sum_{j} (C_{ij})^{-1} C_{j1} N_{1} \ .
    \label{eq:Npop2}
\end{equation}
This suggests the definition of the effective population contribution
coefficient for excitation:
\begin{equation}
    \mathscr{F}_{i1}^{(\mathrm{exc})} = \frac{\sum_j (C_{ij})^{-1} 
    C_{j1}}{N_{\mathrm{e}}} \ .
    \label{eq:eff_cont}
\end{equation}
Hence, the line emissivity can be expressed as
\begin{equation} \varepsilon_{i \rightarrow k} =
    N_{\textrm{e}}\, N_1\, \mathcal{PEC}_{1,i\rightarrow
    k}^{(\mathrm{exc})} \ ,
    \label{eq:emissivity2}
\end{equation}
where the definition for the excitation photon emissivity coefficient
(\PEC) has been used:
\begin{equation}
    \PEC_{1, i\rightarrow k}^{(\mathrm{exc})} \equiv \ 
    \mathscr{F}_{i1}^{(\mathrm{exc})} A_{i\rightarrow k} \ .
    \label{eq:PEC}
\end{equation}

The \PEC\ is a useful intermediate data type, and a more intuitive sense of it
can be obtained by considering its form in the low density limit where
collisional (de-)excitation between excited levels is neglected. Thus,
recalling \eref{eq:cr_elements}, the collisional coupling coefficients between
excited levels become $C_{ij} = A_{j\rightarrow i}\slash N_{\mathrm{e}}$, and
from the ground $C_{i1} =  q^{\mathrm{e}}_{1 \rightarrow i}$. Accordingly, the
low density limit for the excitation \PEC\ is
\begin{equation}
    \PEC^{(\mathrm{exc})}_{1, i \rightarrow k} = 
    \frac{q^{\mathrm{e}}_{1 \rightarrow i} A_{i\rightarrow k}}{\sum_{j < i}
    A_{i\rightarrow j}}\ .
    \label{eq:lowdens3}
\end{equation}
So in the low density limit, the excitation \PEC\ is given by the product of
the EIE rate coefficient from the ground and the branching ratio of the
radiative decay. This reaffirms the assumptions that have been made: the
excited state levels are populated solely by collisional excitation from the
ground and subsequently de-populated by spontaneous emission to any possible
lower level. Therefore, the \PEC\ is an effective quantity for estimating the
diagnostic importance of a transition because it accounts for the population
distribution of levels, a conclusion that equally applies in the more complex,
finite density scenario. 

It is the unsimplified version of the excitation \PEC\ in \eref{eq:PEC} that
will be used by routines in the Atomic Data and Analysis Structure (ADAS)
\cite{adas} for our analysis. These routines use effective collision strengths
produced in the manner described above and stored in the adf04 file format.
Additionally, relativistic effects can cause classically weak, higher order
electric and magnetic radiative transitions to approach similar magnitudes as
the typically dominant dipole ($E1$) transitions; therefore, accurate atomic
population modelling requires inclusion of at least some non-dipole transition
probabilities, $A_{i\rightarrow j}$, for high $z$ ions. {\sc pstgf} only produces $E1$
data derived from the dipole long-range coupling coefficients, so we substituted
$E1$, $E2$/$M1$, and $E3$/$M2$ radiative data from \grasp{} into our final
adf04 file. Comparison with the $A_{i\rightarrow j}$ values in the adf04 file
of Ballance and Griffin revealed that they only include radiative transitions
up to the quadrupole $(E2/M2)$. We include the extra $E3$ data because of the
overlapping selection rules and comparable magnitudes with $M2$.
Further comparison of the radiative data is conducted in the proceeding section
\ref{sub:Structure Data}. 




\section{Results and Discussion} 
\label{sec:Results and Discussion}

\subsection{Structure Data} 
\label{sub:Structure Data}

A portion of our energy level results are summarised in table
\ref{tab:E_levels} along with comparison to other experimental and theoretical
values. Errors relative to the NIST compiled experimental values are given in
brackets for all theoretical calculations. The theoretical results are from the
following calculations: the present \grasp\ and \AS, Ballance and Griffin's
\grasp\ \cite{ballance2007}, and Safronova and Safronova's relativistic
many-body perturbation theory (RMBPT) \cite{safronova2010}. We note that a
recent calculation by Spencer et al \cite{spencer2014} has been omitted from
our detailed comparison to follow.  Although their calculation includes the
important 3d$^9$ core configurations, it uses non-relativistic radial orbitals.
The authors themselves note that their largest error is likely unaccounted
relativistic effects, and and so we restrict detailed comparisons to methods
that use fully or kappa-averaged relativistic radial orbitals. We briefly
comment that our structure results have a similar degree of agreement with
Spencer et al as the other fully relativistic results in their study. 

From a qualitative observation of the errors in table \ref{tab:E_levels}, it is
evident that the Safronova and Safronova theoretical results are closest to the
experimental NIST results.  Moreover, our \grasp\ and Ballance and Griffin's
\grasp\ results appear to be of similar accuracy, while the \AS\ results
perform relatively worst but objectively still quite well. This ordering can be
predicted somewhat since one would not expect the \AS\ calculations that employ
the $\kappa$-averaged Dirac equation to outperform the fully $\kappa$-dependent
Dirac equation used in the other calculations. The \grasp\ values should be
quite similar since they are from the same code but with different CI
expansions, and the Safronova and Safronova values derive from a paper that
focussed exclusively on the atomic structure problem and thus did not need to
balance time and computational resources with a corresponding collision
calculation.

\begin{table}[htbp] \begin{adjustwidth}{-0.5in}{-0.5in} \begingroup
        \fontsize{8pt}{10pt}\selectfont \caption{Lowest 50 energy level
            eigenvalues for \Wff.  All values are in cm$^{-1}$. The bracketed
            values to the right of some theoretical values denote the absolute
            and percent difference from the experimental NIST values,
            respectively. The $jj$-term assignment is strictly for the present
            \grasp\ calculations; equivalence of levels between different
            results is determined on a symmetry ($J\pi$) and energy ($E$)
            mapping. The subscripts have the following meanings. NIST denotes
            the NIST experimental values compiled from various sources
            \cite{kramida2011}; GR denotes the present \grasp\ results; AS
            denotes the present \AS\ results; BG07 denotes the Ballance
            and Griffin results \cite{ballance2007}; and SS10 denotes the
            Safronova and Safronova results \cite{safronova2010}.  }
            \begin{center}
                \begin{tabular}{llllllll} \hline \hline \textbf{$i$} &
                    \textbf{$jj$-term} & \textbf{$J$} &
                    \textbf{$E_{\mathrm{NIST}}$} &
                    \textbf{$E_{\mathrm{GR}}$}
                    &\textbf{$E_{\mathrm{AS}}$}& \textbf{$E_{\mathrm{BG07}}$} &
                    \textbf{$E_{\mathrm{SS10}}$} \\ \hline 1 & 4s$^2$ (1/2,1/2) & 0 &
                    0 & 0 & \multicolumn{1}{l}{0} & 0 & 0 \\ 2 & 4s4p
                    (1/2,1/2)$^{\circ}$ & 0 & 695000 &
                    696599(-1599\textbackslash0.23\%) &
                    \multicolumn{1}{l}{680476(14524\textbackslash2.09\%)} &
                    697338(-2338\textbackslash0.34\%) &
                    696870(-1870\textbackslash0.27\%) \\ 3 & 4s4p
                    (1/2,1/2)$^{\circ}$ & 1 & 752560 &
                    754900(-2340\textbackslash0.31\%) &
                    \multicolumn{1}{l}{738077(14483\textbackslash1.92\%)} &
                    756118(-3558\textbackslash0.47\%) &
                    752290(270\textbackslash0.04\%) \\ 4 & 4s4p
                    (1/2,3/2)$^{\circ}$ & 2 & 1494400 &
                    1510410(-16010\textbackslash1.07\%) &
                    \multicolumn{1}{l}{1500353(-5953\textbackslash0.40\%)} &
                    1511424(-17024\textbackslash1.14\%) &
                    1505330(-10930\textbackslash0.73\%) \\ 5 & 4p$^2$ (1/2,1/2)
                    & 0 & 1588000 & 1610234(-22234\textbackslash1.40\%) &
                    \multicolumn{1}{l}{1598341(-10341\textbackslash0.65\%)} &
                    1603286(-15286\textbackslash0.96\%) &
                    1589470(-1470\textbackslash0.09\%) \\ 6 & 4s4p
                    (1/2,3/2)$^{\circ}$ & 1 & 1641230 &
                    1654698(-13468\textbackslash0.82\%) &
                    \multicolumn{1}{l}{1645076(-3846\textbackslash0.23\%)} &
                    1657295(-16065\textbackslash0.98\%) &
                    1641860(-630\textbackslash0.04\%) \\ 7 & 4p$^2$ (1/2,3/2) &
                    1 & 2345700 & 2370326(-24626\textbackslash1.05\%) &
                    \multicolumn{1}{l}{2367366(-21666\textbackslash0.92\%)} &
                    2364982(-19282\textbackslash0.82\%) &
                    2347790(-2090\textbackslash0.09\%) \\ 8 & 4p$^2$ (1/2,3/2)
                    & 2 & 2362700 & 2380945(-18245\textbackslash0.77\%) &
                    \multicolumn{1}{l}{2380127(-17427\textbackslash0.74\%)} &
                    2375598(-12898\textbackslash0.55\%) &
                    2359810(2890\textbackslash0.12\%) \\ 9 & 4s4d (1/2,3/2) & 1
                    & 2782700 & 2807138(-24438\textbackslash0.88\%) &
                    \multicolumn{1}{l}{2826740(-44040\textbackslash1.58\%)} &
                    2801178(-18478\textbackslash0.66\%) &
                    2781700(1000\textbackslash0.04\%) \\ 10 & 4s4d (1/2,3/2) &
                    2 & 2809500 & 2835916(-26416\textbackslash0.94\%) &
                    \multicolumn{1}{l}{2854715(-45215\textbackslash1.61\%)} &
                    2829810(-20310\textbackslash0.72\%) &
                    2809010(490\textbackslash0.02\%) \\ 11 & 4s4d (1/2,5/2) & 3
                    & 2943800 & 2980289(-36489\textbackslash1.24\%) &
                    \multicolumn{1}{l}{3007602(-63802\textbackslash2.17\%)} &
                    2974581(-30781\textbackslash1.05\%) &
                    2952430(-8630\textbackslash0.29\%) \\ 12 & 4s4d (1/2,5/2) &
                    2 & 2988500 & 3025731(-37231\textbackslash1.25\%) &
                    \multicolumn{1}{l}{3047061(-58561\textbackslash1.96\%)} &
                    3019918(-31418\textbackslash1.05\%) &
                    2997790(-9290\textbackslash0.31\%) \\ 13 & 4p$^2$ (3/2,3/2)
                    & 2 & 3210900 & 3244954(-34054\textbackslash1.06\%) &
                    \multicolumn{1}{l}{3254573(-43673\textbackslash1.36\%)} &
                    3239406(-28506\textbackslash0.89\%) &
                    3211110(-210\textbackslash0.01\%) \\ 14 & 4p$^2$ (3/2,3/2)
                    & 0 & 3249000 & 3283304(-34304\textbackslash1.06\%) &
                    \multicolumn{1}{l}{3288983(-39983\textbackslash1.23\%)} &
                    3277012(-28012\textbackslash0.86\%) &
                    3251480(-2480\textbackslash0.08\%) \\ 15 & 4p4d
                    (1/2,3/2)$^{\circ}$ & 2 &  & 3542869 & 3548176 & 3536793 &
                    3516410 \\ 16 & 4p4d (1/2,3/2)$^{\circ}$ & 1 &  & 3686507 &
                    3685971 & 3679726 & 3649830 \\ 17 & 4p4d
                    (1/2,5/2)$^{\circ}$ & 3 &  & 3793159 & 3802977 & 3786985 &
                    3759910 \\ 18 & 4p4d (1/2,5/2)$^{\circ}$ & 2 &  & 3795417 &
                    3804873 & 3789273 & 3760590 \\ 19 & 4s4f
                    (1/2,5/2)$^{\circ}$ & 3 &  & 4296920 & 4306386 & 4292056 &
                    4268490 \\ 20 & 4s4f (1/2,7/2)$^{\circ}$ & 2 &  & 4324408 &
                    4333915 & 4319207 & 4293610 \\ 21 & 4s4f
                    (1/2,5/2)$^{\circ}$ & 4 &  & 4354514 & 4375712 & 4349717 &
                    4324560 \\ 22 & 4s4f (1/2,5/2)$^{\circ}$ & 3 &  & 4381359 &
                    4401322 & 4376049 & 4347880 \\ 23 & 4p4d
                    (3/2,3/2)$^{\circ}$ & 2 & 4383000 &
                    4422045(-39045\textbackslash0.89\%) &
                    \multicolumn{1}{l}{4431516(-48516\textbackslash1.11\%)} &
                    4416368(-33368\textbackslash0.76\%) &
                    4385180(-2180\textbackslash0.05\%) \\ 24 & 4p4d
                    (3/2,3/2)$^{\circ}$ & 0 &  & 4443019 & 4451669 & 4437256 &
                    4406260 \\ 25 & 4p4d (3/2,3/2)$^{\circ}$ & 1 &  & 4453869 &
                    4463503 & 4448129 & 4415630 \\ 26 & 4p4d
                    (3/2,3/2)$^{\circ}$ & 3 & 4458000 &
                    4501932(-43932\textbackslash0.99\%) &
                    \multicolumn{1}{l}{4512851(-54851\textbackslash1.23\%)} &
                    4495374(-37374\textbackslash0.84\%) &
                    4460510(-2510\textbackslash0.06\%) \\ 27 & 4p4d
                    (3/2,5/2)$^{\circ}$ & 4 & 4505300 &
                    4547619(-42319\textbackslash0.94\%) &
                    \multicolumn{1}{l}{4564787(-59487\textbackslash1.32\%)} &
                    4541971(-36671\textbackslash0.81\%) &
                    4511020(-5720\textbackslash0.13\%) \\ 28 & 4p4d
                    (3/2,5/2)$^{\circ}$ & 2 &  & 4587583 & 4604286 & 4582203 &
                    4549230 \\ 29 & 4p4d (3/2,5/2)$^{\circ}$ & 1 &  & 4711801 &
                    4729592 & 4705746 & 4667050 \\ 30 & 4p4d
                    (3/2,5/2)$^{\circ}$ & 3 & 4667000 &
                    4720344(-53344\textbackslash1.14\%) &
                    \multicolumn{1}{l}{4738811(-71811\textbackslash1.54\%)} &
                    4712765(-45765\textbackslash0.98\%) &
                    4669890(-2890\textbackslash0.06\%) \\ 31 & 4p4f (1/2,5/2) &
                    3 &  & 5106504 & 5099115 & 5101065 & 5069120 \\ 32 & 4p4f
                    (1/2,5/2) & 2 &  & 5149812 & 5139452 & 5144560 & 5110970 \\
                    33 & 4p4f (1/2,7/2) & 3 &  & 5174655 & 5178086 & 5169893 &
                    5135570 \\ 34 & 4p4f (1/2,7/2) & 4 &  & 5175709 & 5179078 &
                    5169835 & 5136020 \\ 35 & 4d$^2$ (3/2,3/2) & 2 &  & 5671068
                    & 5684603 & 5662259 & 5621680 \\ 36 & 4d$^2$ (3/2,3/2) & 0
                    &  & 5746101 & 5759234 & 5732092 & 5690100 \\ 37 & 4d$^2$
                    (3/2,5/2) & 3 &  & 5808133 & 5826269 & 5801275 & 5762150 \\
                    38 & 4d$^2$ (3/2,5/2) & 4 &  & 5816599 & 5831429 & 5810323
                    & 5772640 \\ 39 & 4d$^2$ (3/2,5/2) & 2 &  & 5843017 &
                    5861428 & 5834560 & 5794100 \\ 40 & 4d$^2$ (3/2,5/2) & 1 &
                    & 5877633 & 5898701 & 5866642 & 5823700 \\ 41 & 4p4f
                    (3/2,7/2) & 4 &  & 5917488 & 5934565 & 5912767 & 5876050 \\
                    42 & 4p4f (3/2,5/2) & 3 &  & 5927978 & 5932078 & 5922956 &
                    5884140 \\ 43 & 4p4f (3/2,5/2) & 2 &  & 5957126 & 5961349 &
                    5951431 & 5910810 \\ 44 & 4p4f (3/2,7/2) & 5 &  & 5970835 &
                    5983185 & 5965829 & 5926610 \\ 45 & 4p4f (3/2,5/2) & 1 &  &
                    5971049 & 5970842 & 5966743 & 5927040 \\ 46 & 4p4f
                    (3/2,7/2) & 3 &  & 5986398 & 5999033 & 5981863 & 5941010 \\
                    47 & 4p4f (3/2,5/2) & 4 &  & 5991059 & 6004943 & 5983469 &
                    5938830 \\ 48 & 4d$^2$ (3/2,3/2) & 2 &  & 6007925 & 6028552
                    & 6000992 & 5958400 \\ 49 & 4p4f (3/2,7/2) & 2 &  & 6114914
                    & 6137006 & 6105042 & 6055560 \\ 50 & 4d$^2$ (5/2,5/2) & 4
                    &  & 6137752 & 6158685 & 6126041 & 6072960 \\ 
                \hline
            \end{tabular} \end{center} \label{tab:E_levels}
    \endgroup \end{adjustwidth} \end{table}

\begin{figure}[htbp]
    \centering
    \includegraphics[width=\linewidth]{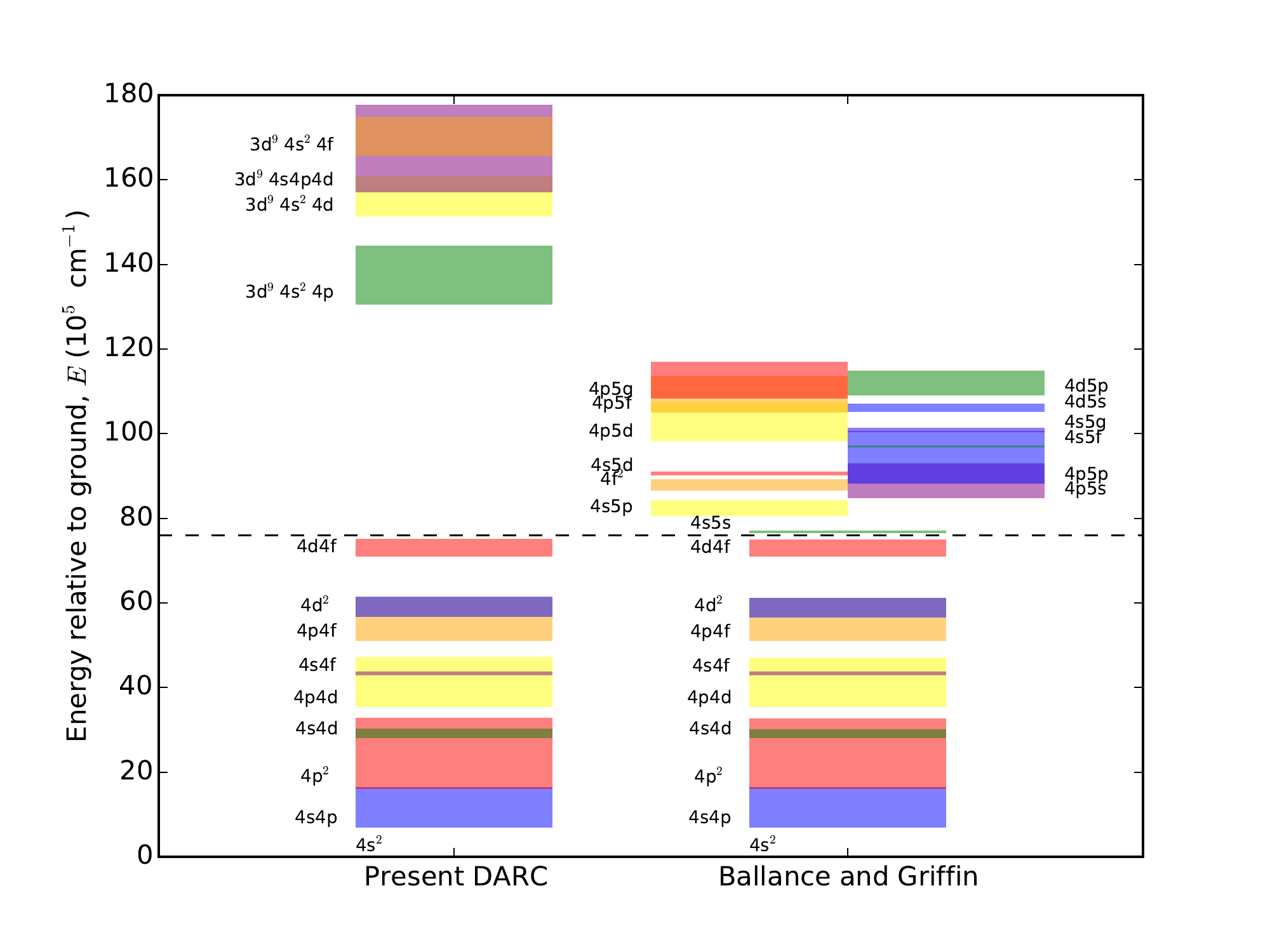}
    \caption{Energy ranges of the configurations included in the present \darc{}
    calculations and the Ballance and Griffin calculations. Non-relativistic
    configuration specifications are used for brevity with the understanding
    they encompass multiple relativistic sub-configurations. The energy ranges are
    determined by assigning each $jj$-coupled level to the corresponding
    configuration which contributes the dominant component the level's state
vector. This method can be ambiguous in cases where strong configuration mixing
is present.}
    \label{fig:config_energies}
\end{figure}

Because we will be comparing extensively with the Ballance and Griffin results,
it is important to obtain an overall concept of how the energy levels compare
between the two calculations, something difficult to grasp from raw data
tables.  Accordingly, figure \ref{fig:config_energies} provides an illustrative
graphic of the energy ranges of the configurations included in the two
calculations.  Below approximately $8 \times 10^6$ cm$^{-1}$, the configuration
energy ranges visually match to a small degree of error. This is quantitatively
substantiated by the proximity of the energy levels in table \ref{tab:E_levels}
and a mean percent difference of 0.13\% for all intersecting levels. However,
above this threshold, the energy ranges are completely discrepant owing to the
differences in the CI expansions. In our calculations (left), there is an
energy gap between the first open 3d-subshell configuration (3d$^9$4s$^2$4p)
and the highest closed 3d-subshell configuration (4d4f). On the other hand, the
3d$^{10}$4$l$5$l^{\prime}$ configurations, which Ballance and Griffin include,
coincidently and neatly fill this energy gap. The implications of this gross
difference in energy level distribution will be investigated throughout the
remainder of the paper, especially in relation to the CC expansion and effect
upon the collision data. 

Additionally, a sample of the radiative data from our \grasp\ structure results
is presented in table \ref{tab:rad_data}. Apart from wavelengths, negligible
experimental radiative data is available, and so only theoretical results are
supplied for comparison. The theoretical results are from the same calculations
as in the energy level table \ref{tab:E_levels}, excepting the addition of
Fournier's \emph{ab initio} calculations \cite{fournier1998} and the omission
of our \AS{} results for brevity. The Fournier $gf$ values for the 212--1 and
290--1 transitions are discrepant because the 3d$^9$4s4p4d configuration was
not included in that calculation, and as demonstrated in section \ref{sub:CI
and Structure Determination}, the 3d$^9$4s4p4d configuration mixes heavily and
greatly changes the radiative data of these 3d-subshell transitions.
Consequently, comparison of these transitions with calculations that do not
include this configuration are not meaningful.  Otherwise, the Fournier $gf$
values tend to agree well with our corresponding \grasp\ results, except for
the rather weak transitions 129--6 and 73--10 that differ by about a factor of
three.

The Ballance and Griffin \grasp\ results also appear to be in close agreement
with our \grasp\ results in this sample, except in instances where the
magnitude of the $gf$ value is small or the velocity to length ratios are not
close to unity. In both cases, this is to be expected when comparing
calculations with different CI expansions. A full scope but necessarily more
coarse comparison with our results was conducted using scatter plots analogous
to those in figure \ref{fig:lin_plots}. Neither the dipole line strengths,
$S_{ik}$, nor the radiative transition probabilities up to quadrupole order
revealed any systematic differences between the calculations, and 73\% of the
values agree within 20\% relative error of each other, meaning there is
reasonable accord overall. The dipole line strengths are directly proportional
to the infinite energy limits of the corresponding EIE collision strength, and
so this information will be relevant for the analysis of the collision data in
section \ref{sub:Collision Data}. 

On the other hand, the Safronova and Safronova results exhibit a binary
behaviour: they either agree well with the present results or disagree by a few orders
of magnitude. Based on the energy level values quoted by Safronova and
Safronova, we can say with a high degree of certainty that this disagreement is
not due to a level mismatching by us; however, we did observe significant
differences in the wavelength values for these conflicting transitions. Upon
further investigation, the wavelengths given by Safronova and Safronova do not
agree with their own energy level values. Thus, we suspect that there has been
a labelling error in their work. To confirm this hypothesis, we investigated
further with \AS\ to provide a corroborative third party result. We already had
the relevant results from using the CI expansion in section 
\ref{sub:CI and Structure Determination}, and an additional
run was conducted using the CI from the Safronova and Safronova work. In both
cases, the \AS\ results agreed with the present \grasp\ results, supporting the
validity of the present work and pointing to a labelling error in the Safronova and
Safronova results.

\begin{table}[htbp]
\begin{adjustwidth}{-0.5in}{-0.5in} 
    \begingroup 
    \fontsize{8pt}{10pt}\selectfont
    \caption{Radiative data: weighted oscillator strength ($gf$) and wavelength
        ($\lambda$) values for \Wff. GR denotes the
        present results generated using \grasp; F98 denotes the results from
        Fournier \cite{fournier1998}; BG07 denotes the results from
        Ballance and Griffin \cite{ballance2007}; and SS10 denotes the results
        from Safronova and Safronova \cite{safronova2010}. The \AS\
        results are not presented in the interest of brevity. The level
        specifications are for the present results, and mapping of levels
        between the different calculations was determined by matching symmetry
        ($J\pi$) and energy ($E$), as in the case of the energy level
        table. Conversion from $A_{ki}$ values to $gf$ values for the BG07 data
        was necessary for comparision, and we used their calculated energies to
        do so. For compactness, $\star = $
        (3d$^9$($^2$D$_{5/2}$)4s$_{1/2}$)$^{\circ}_2$4p$_{3/2}$. All results
    are in the length gauge, and $v/l$ denotes the ratio of the velocity gauge
    to the length gauge. Values presented in the format
    $\mathrm{X.XXX}\pm\mathrm{YY}$ represent
scientific notation in base 10: $\mathrm{X.XXX}\times 10^{\pm \mathrm{YY}}$}.
\begin{center}
\begin{tabular}{rrlrrrrrrrrr}

    \hline \hline
\multicolumn{1}{l}{$i$} & \multicolumn{1}{l}{$k$} & $jj$-coupled CSF of $k$ &
\multicolumn{1}{l}{$J_i$} & \multicolumn{1}{l}{$J_k$} &
\multicolumn{1}{l}{$gf_{\textrm{GR}}$} &
\multicolumn{1}{l}{$v/l_{\textrm{GR}}$} &
\multicolumn{1}{l}{$gf_{\textrm{BG07}}$} &
\multicolumn{1}{l}{$v/l_{\textrm{BG07}}$} &
\multicolumn{1}{l}{$gf_{\textrm{F98}}$} &
\multicolumn{1}{l}{$gf_{\textrm{SS10}}$} & \multicolumn{1}{l}{$\lambda_{\textrm{GR}}$ (\AA)}
\\ \hline
1 & 295 &  ($\star$)$^{\circ}_{7/2}$4d$_{5/2}$ (7/2,5/2)$^{\circ}$ & 0 & 1 & 9.028$-$01 & 0.89 & $-$ & $-$ & $-$ & $-$ & 5.7330 \\ 
1 & 290 &  3d$^9$($^2$D$_{3/2}$)4s$^2$4f (3/2,5/2)$^{\circ}$ & 0 & 1 & 1.610$+$00 & 0.90 & $-$ & $-$ & 5.844$+$00 & $-$ & 5.7438 \\ 
1 & 275 &  ($\star$)$^{\circ}_{1/2}$4d$_{3/2}$ (1/2,3/2)$^{\circ}$ & 0 & 1 & 1.894$+$00 & 0.91 & $-$ & $-$ & $-$ & $-$ & 5.7917 \\ 
1 & 212 &  ($\star$)$^{\circ}_{3/2}$4d$_{5/2}$ (3/2,5/2)$^{\circ}$ & 0 & 1 & 3.820$-$01 & 0.89 & $-$ & $-$ & 1.954$+$00 & $-$ & 5.9485 \\ 
1 & 208 &  ($\star$)$^{\circ}_{5/2}$4d$_{3/2}$ (5/2,3/2)$^{\circ}$ & 0 & 1 & 4.201$-$01 & 0.91 & $-$ & $-$ & $-$ & $-$ & 5.9616 \\ 
1 & 207 &  ($\star$)$^{\circ}_{3/2}$4d$_{5/2}$ (3/2,5/2)$^{\circ}$ & 0 & 1 & 4.923$-$01 & 0.92 & $-$ & $-$ & $-$ & $-$ & 5.9655 \\ 
1 & 81 &  3d$^9$($^2$D$_{3/2}$)4s$^2$4p (3/2,3/2)$^{\circ}$ & 0 & 1 & 2.912$-$02 & 0.91 & $-$ & $-$ & 2.800$-$02 & $-$ & 6.9483 \\ 
6 & 129 &  3d$^9$($^2$D$_{3/2}$)4s$^2$4d (3/2,3/2) & 1 & 0 & 5.017$-$04 & 0.00 & $-$ & $-$ & 1.292$-$03 & $-$ & 6.9367 \\ 
1 & 78 &  3d$^9$($^2$D$_{5/2}$)4s$^2$4p (5/2,3/2)$^{\circ}$ & 0 & 1 & 2.562$-$01 & 0.91 & $-$ & $-$ & 2.379$-$01 & $-$ & 7.2056 \\ 
1 & 75 &  3d$^9$($^2$D$_{3/2}$)4s$^2$4p (3/2,1/2)$^{\circ}$ & 0 & 1 & 1.519$-$01 & 0.91 & $-$ & $-$ & 1.412$-$01 & $-$ & 7.3524 \\ 
4 & 83 &  3d$^9$($^2$D$_{3/2}$)4s$^2$4p (3/2,3/2)$^{\circ}$ & 2 & 2 & 1.580$-$04 & 0.90 & $-$ & $-$ & 1.488$-$04 & $-$ & 7.7453 \\ 
4 & 82 &  3d$^9$($^2$D$_{3/2}$)4s$^2$4p (3/2,3/2)$^{\circ}$ & 2 & 3 & 1.303$-$04 & 0.01 & $-$ & $-$ & 1.237$-$04 & $-$ & 7.7580 \\ 
2 & 74 &  3d$^9$($^2$D$_{3/2}$)4s$^2$4p (3/2,1/2)$^{\circ}$ & 0 & 2 & 9.193$-$05 & 2.20 & $-$ & $-$ & 8.710$-$05 & $-$ & 7.7670 \\ 
3 & 74 &  3d$^9$($^2$D$_{3/2}$)4s$^2$4p (3/2,1/2)$^{\circ}$ & 1 & 2 & 1.301$-$04 & 8.70 & $-$ & $-$ & 1.294$-$04 & $-$ & 7.8015 \\ 
6 & 82 &  3d$^9$($^2$D$_{3/2}$)4s$^2$4p (3/2,3/2)$^{\circ}$ & 1 & 3 & 1.875$-$04 & 0.08 & $-$ & $-$ & 1.738$-$04 & $-$ & 7.8462 \\ 
4 & 79 &  3d$^9$($^2$D$_{5/2}$)4s$^2$4p (5/2,3/2)$^{\circ}$ & 2 & 3 & 1.999$-$04 & 0.01 & $-$ & $-$ & 1.839$-$04 & $-$ & 8.0730 \\ 
4 & 77 &  3d$^9$($^2$D$_{5/2}$)4s$^2$4p (5/2,3/2)$^{\circ}$ & 2 & 2 & 7.144$-$05 & 0.91 & $-$ & $-$ & 6.886$-$05 & $-$ & 8.0880 \\ 
4 & 76 &  3d$^9$($^2$D$_{5/2}$)4s$^2$4p (5/2,3/2)$^{\circ}$ & 2 & 4 & 4.136$-$04 & 0.01 & $-$ & $-$ & 3.961$-$04 & $-$ & 8.0991 \\ 
2 & 72 &  3d$^9$($^2$D$_{5/2}$)4s$^2$4p (5/2,1/2)$^{\circ}$ & 0 & 2 & 1.379$-$04 & 0.88 & $-$ & $-$ & 1.313$-$04 & $-$ & 8.0998 \\ 
3 & 73 &  3d$^9$($^2$D$_{5/2}$)4s$^2$4p (5/2,1/2)$^{\circ}$ & 1 & 3 & 3.229$-$04 & 0.01 & $-$ & $-$ & 2.998$-$04 & $-$ & 8.1327 \\ 
3 & 72 &  3d$^9$($^2$D$_{5/2}$)4s$^2$4p (5/2,1/2)$^{\circ}$ & 1 & 2 & 8.764$-$05 & 3.00 & $-$ & $-$ & 8.684$-$05 & $-$ & 8.1380 \\ 
6 & 77 &  3d$^9$($^2$D$_{5/2}$)4s$^2$4p (5/2,3/2)$^{\circ}$ & 1 & 2 & 1.469$-$04 & 0.04 & $-$ & $-$ & 1.494$-$04 & $-$ & 8.1840 \\ 
11 & 76 &  3d$^9$($^2$D$_{5/2}$)4s$^2$4p (5/2,3/2)$^{\circ}$ & 3 & 4 & 1.940$-$04 & 2.00 & $-$ & $-$ & 3.014$-$04 & $-$ & 9.1878 \\ 
10 & 73 &  3d$^9$($^2$D$_{5/2}$)4s$^2$4p (5/2,1/2)$^{\circ}$ & 2 & 3 & 3.580$-$05 & 2.90 & $-$ & $-$ & 1.389$-$04 & $-$ & 9.7800 \\ 
3 & 12 &  4s4d (1/2,5/2) & 1 & 2 & 8.387$-$02 & 1.00 & 7.694$-$02 & $-$ & 8.768$-$02 & 7.500$-$02 & 44.2929 \\ 
3 & 10 &  4s4d (1/2,3/2) & 1 & 2 & 1.775$+$00 & 1.00 & 1.795$+$00 & $-$ & 1.776$+$00 & 1.689$+$00 & 48.2882 \\ 
1 & 6 &  4s4p (1/2,3/2)$^{\circ}$ & 0 & 1 & 1.095$+$00 & 0.83 & 1.139$+$00 & 0.99 & 1.099$+$00 & 1.060$+$00 & 60.6907 \\ 
3 & 8 &  4p$^2$ (1/2,3/2) & 1 & 2 & 7.290$-$01 & 0.99 & 7.460$-$01 & $-$ & 7.256$-$01 & $-$ & 61.6827 \\ 
6 & 13 &  4p$^2$ (3/2,3/2) & 1 & 2 & 2.351$+$00 & 1.00 & 2.393$+$00 & $-$ & 2.404$+$00 & 2.244$+$00 & 63.0756 \\ 
4 & 12 &  4s4d (1/2,5/2) & 2 & 2 & 6.591$-$01 & 1.00 & 6.753$-$01 & $-$ & 6.882$-$01 & 6.350$-$01 & 66.2383 \\ 
6 & 12 &  4s4d (1/2,5/2) & 1 & 2 & 4.271$-$01 & 1.00 & 4.199$-$01 & $-$ & 3.878$-$01 & $-$ & 73.2493 \\ 
1 & 3 &  4s4p (1/2,1/2)$^{\circ}$ & 0 & 1 & 1.364$-$01 & 0.59 & 1.415$-$01 & 1.00 & 1.376$-$01 & 1.320$-$01 & 132.4223 \\ 
3 & 4 &  4s4p (1/2,3/2)$^{\circ}$ & 1 & 2 & 5.643$-$05 & 1.00 & 5.873$-$05 & $-$ & 5.637$-$05 & $-$ & 133.6916 \\ 
1 & 16 & 4p4d (1/2,3/2)$^{\circ}$ & 0 & 1 & 2.185$-$04 & 1.50 & 1.484$-$04 & 0.99 & $-$ & $-$ & 27.1909 \\ 
1 & 29 & 4p4d (3/2,5/2)$^{\circ}$ & 0 & 1 & 1.598$-$04 & 0.95 & 3.392$-$04 & 1.10 & $-$ & $-$ & 21.3138 \\ 
1 & 59 & 4d4f (3/2,5/2)$^{\circ}$ & 0 & 1 & 3.357$-$05 & 0.01 & 4.694$-$05 & 0.91 & $-$ & $-$ & 13.8487 \\ 
1 & 71 & 4d4f (5/2,7/2)$^{\circ}$ & 0 & 1 & 1.885$-$04 & 0.08 & 2.402$-$04 & 1.00 & $-$ & $-$ & 13.3697 \\ 
2 & 7 & 4p$^2$ (1/2,3/2) & 0 & 1 & 5.135$-$01 & 1.00 & 5.191$-$01 & 0.99 & $-$ & $-$ & 59.9089 \\ 
2 & 9 & 4s4d (1/2,3/2) & 0 & 1 & 6.148$-$01 & 1.00 & 6.249$-$01 & 1.00 & $-$ & $-$ & 47.6077 \\ 
2 & 40 & 4d$^2$ (3/2,5/2) & 0 & 1 & 3.898$-$05 & 0.81 & 5.350$-$05 & 0.79 & $-$ & $-$ & 19.3960 \\ 
2 & 45 & 4p4f (3/2,5/2) & 0 & 1 & 6.580$-$05 & 1.10 & 9.251$-$05 & 1.20 & $-$ & $-$ & 19.0480 \\ 
8 & 19 & 4s4f (1/2,5/2)$^{\circ}$ & 2 & 3 & 6.398$-$01 & 1.00 & 6.714$-$01 & $-$ & $-$ & $-$ & 52.5430 \\ 
4 & 11 & 4s4d (1/2,5/2) & 2 & 3 & 1.860$+$00 & 1.00 & 1.887$+$00 & $-$ & $-$ & $-$ & 68.3293 \\ 
75 & 129 & 3d$^9$($^2$D$_{3/2}$)4s$^2$4d (3/2,3/2) & 1 & 0 & 2.243$-$01 & 0.87 & $-$ & $-$ & $-$ & $-$ & 40.5992 \\ 
20 & 45 & 4d$^2$ (3/2,5/2) & 2 & 1 & 7.047$-$01 & 0.87 & $-$ & $-$ & $-$ & 9.000$-$01 & 60.9793 \\ 
7 & 25 & 4p4d (3/2,3/2)$^{\circ}$ & 1 & 1 & 7.485$-$01 & 1.00 & $-$ & $-$ & $-$ & 7.140$-$01 & 48.2905 \\ 
8 & 28 & 4p4d (3/2,5/2)$^{\circ}$ & 2 & 2 & 5.357$-$03 & 1.00 & $-$ & $-$ & $-$ & 8.350$-$01 & 45.6454 \\ 
8 & 26 & 4p4d (3/2,3/2)$^{\circ}$ & 2 & 3 & 1.071$+$00 & 1.00 & $-$ & $-$ & $-$ & 6.550$-$01 & 47.4473 \\ 
10 & 17 & 4s4f (1/2,5/2)$^{\circ}$ & 2 & 3 & 3.092$-$02 & 0.97 & $-$ & $-$ & $-$ & 2.490$+$00 & 104.8515 \\ 
17 & 38 & 4p4f (3/2,5/2) & 3 & 4 & 1.973$+$00 & 1.00 & $-$ & $-$ & $-$ & 4.389$+$00 & 49.7191 \\ 
29 & 39 & 4p4f (3/2,7/2) & 1 & 2 & 4.511$-$02 & 0.94 & $-$ & $-$ & $-$ & 1.614$+$00 & 88.8178 \\ 
\hline
\end{tabular}
\end{center}
\label{tab:rad_data}
\endgroup
\end{adjustwidth}
\end{table}


\subsection{Collision Data} 
\label{sub:Collision Data}

\begin{figure}[htpb] \begin{adjustwidth}{-0.5in}{-0.5in}
        \begin{subfigure}{0.5\linewidth} \begin{center}
                \includegraphics[width=\linewidth]{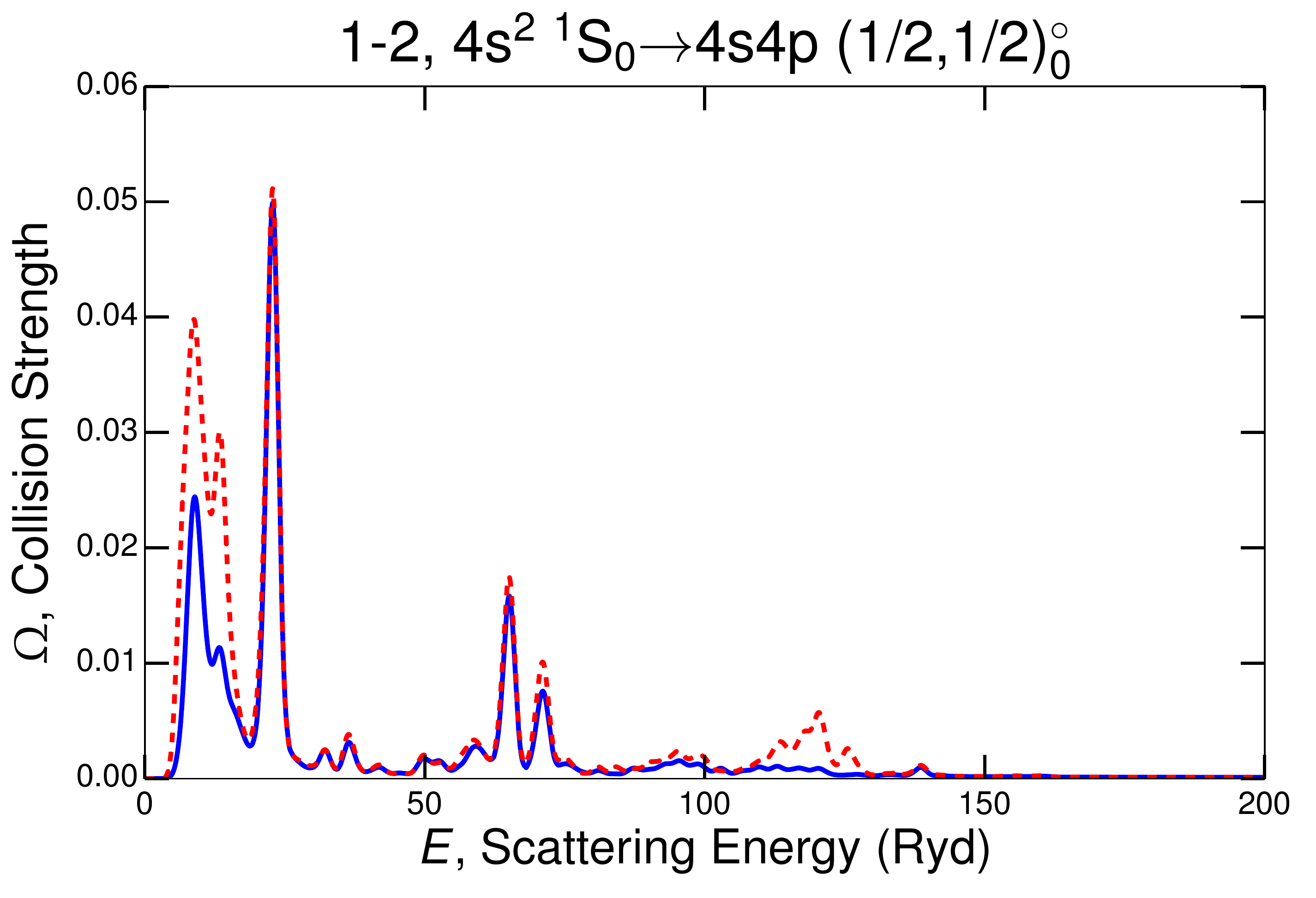}
            \end{center} \vspace{-0.75\baselineskip} \caption{}
            \label{fig:BGom1-2} \end{subfigure}
        \begin{subfigure}{0.5\linewidth} \begin{center}
                \includegraphics[width=\linewidth]{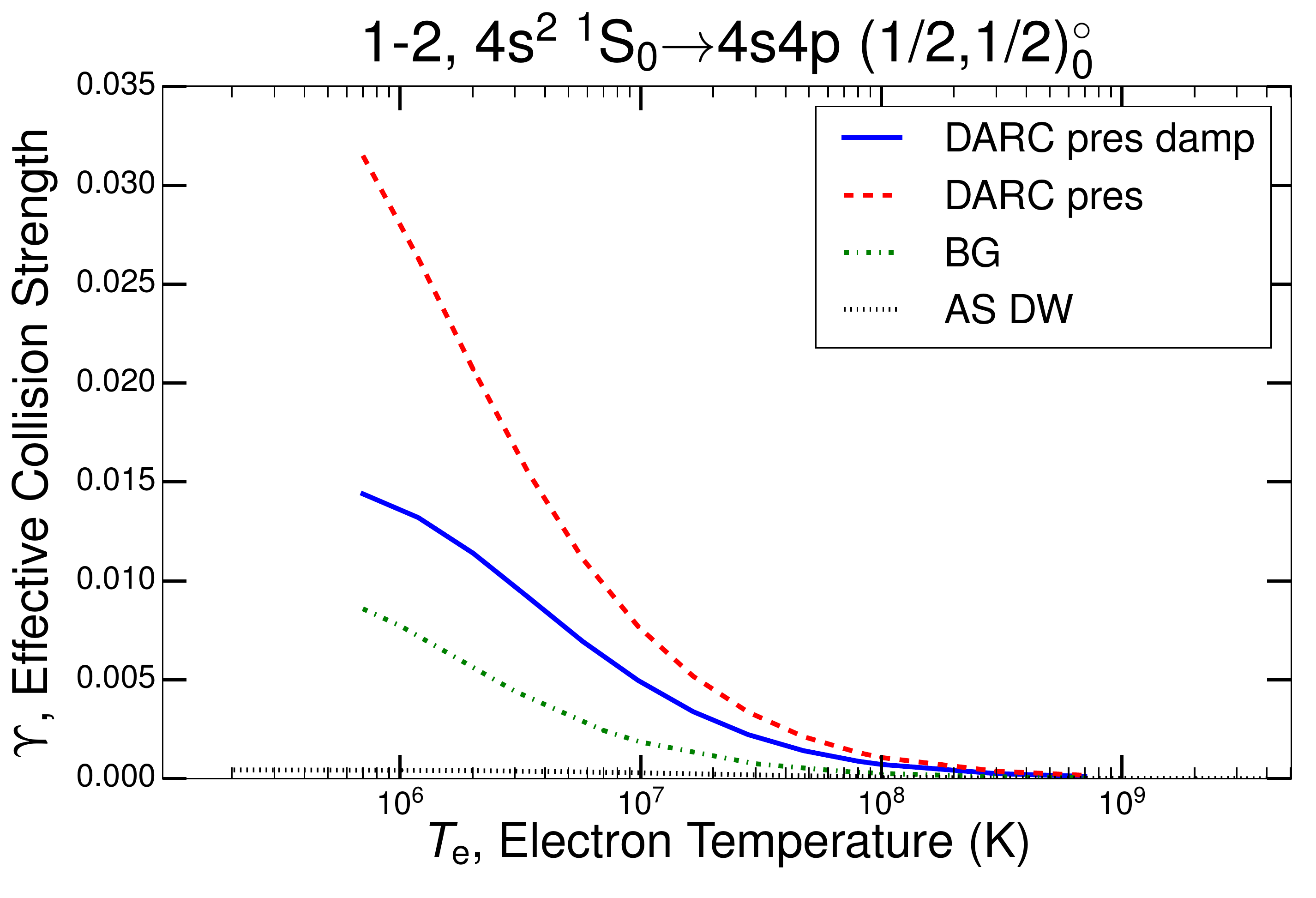}
            \end{center} \vspace{-0.75\baselineskip} \caption{}
            \label{fig:BGups1-2} \end{subfigure} \\
        \begin{subfigure}{0.5\linewidth} \begin{center}
                \includegraphics[width=\linewidth]{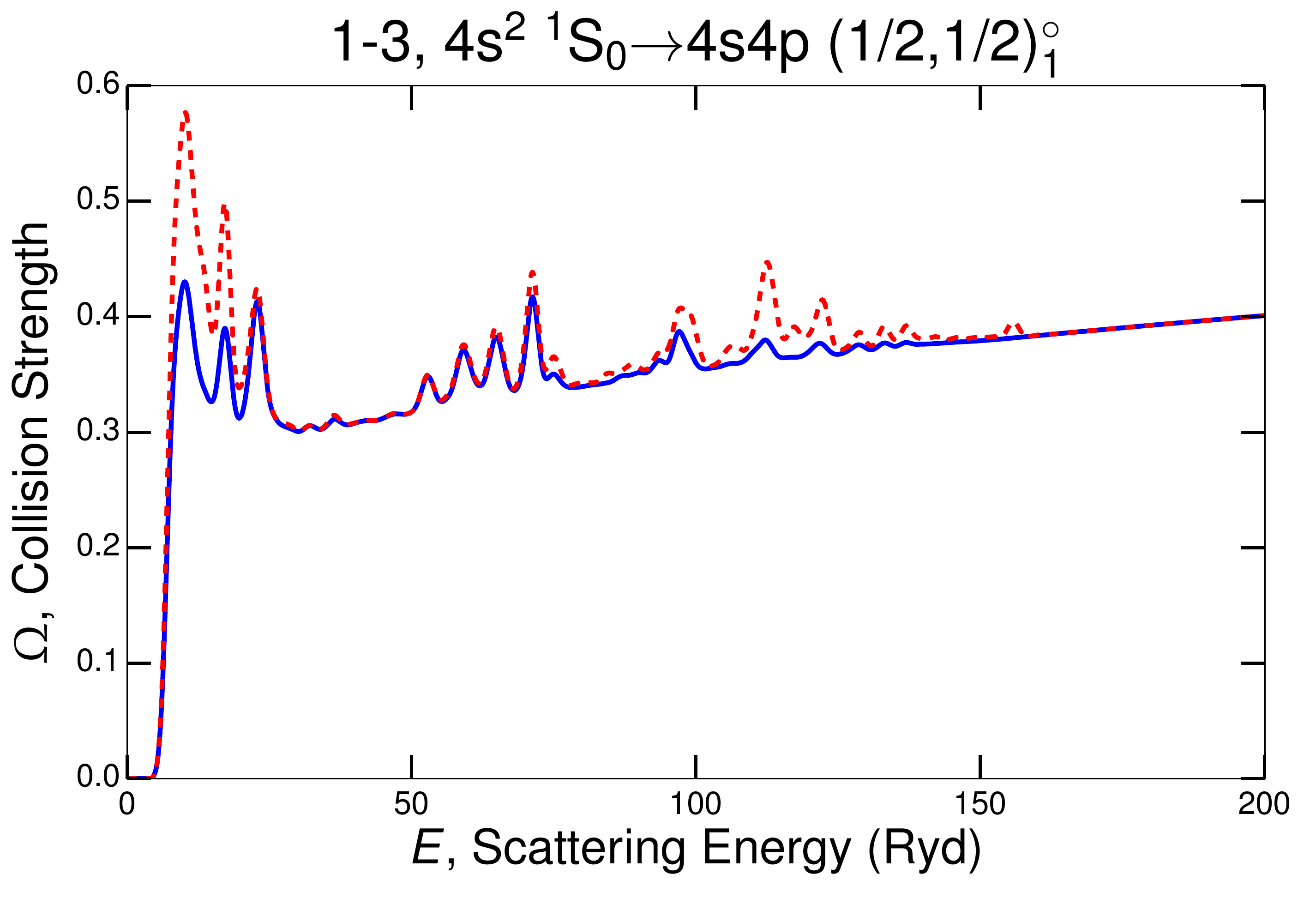}
            \end{center} \vspace{-0.75\baselineskip} \caption{}
            \label{fig:BGom1-3} \end{subfigure}
        \begin{subfigure}{0.5\linewidth} \begin{center}
                \includegraphics[width=\linewidth]{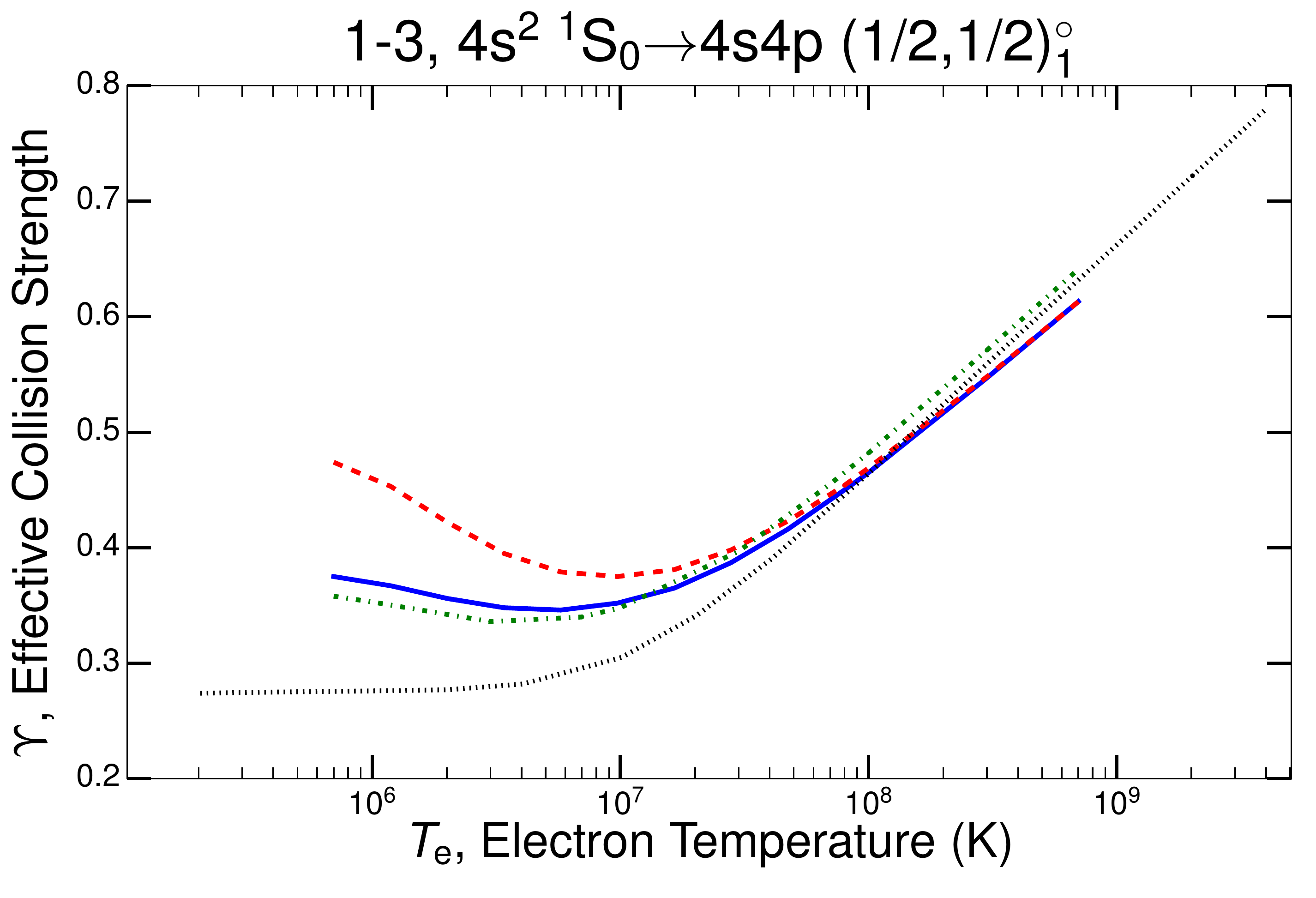}
            \end{center} \vspace{-0.75\baselineskip} \caption{}
            \label{fig:BGups1-3} \end{subfigure} \\
        \begin{subfigure}{0.5\linewidth} \begin{center}
                \includegraphics[width=\linewidth]{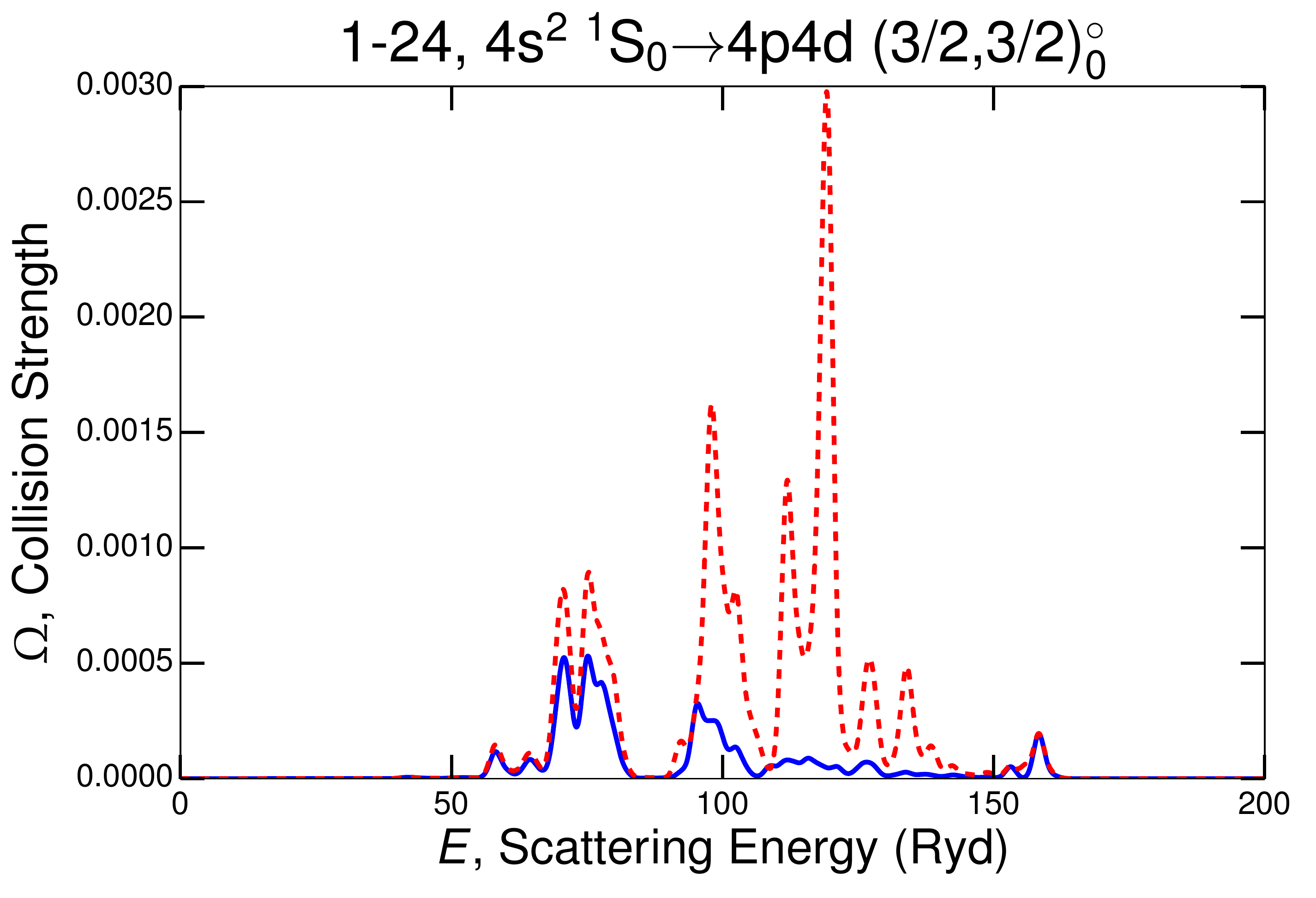}
            \end{center} \vspace{-0.75\baselineskip} \caption{}
            \label{fig:BGom1-24} \end{subfigure}
        \begin{subfigure}{0.5\linewidth} \begin{center}
                \includegraphics[width=\linewidth]{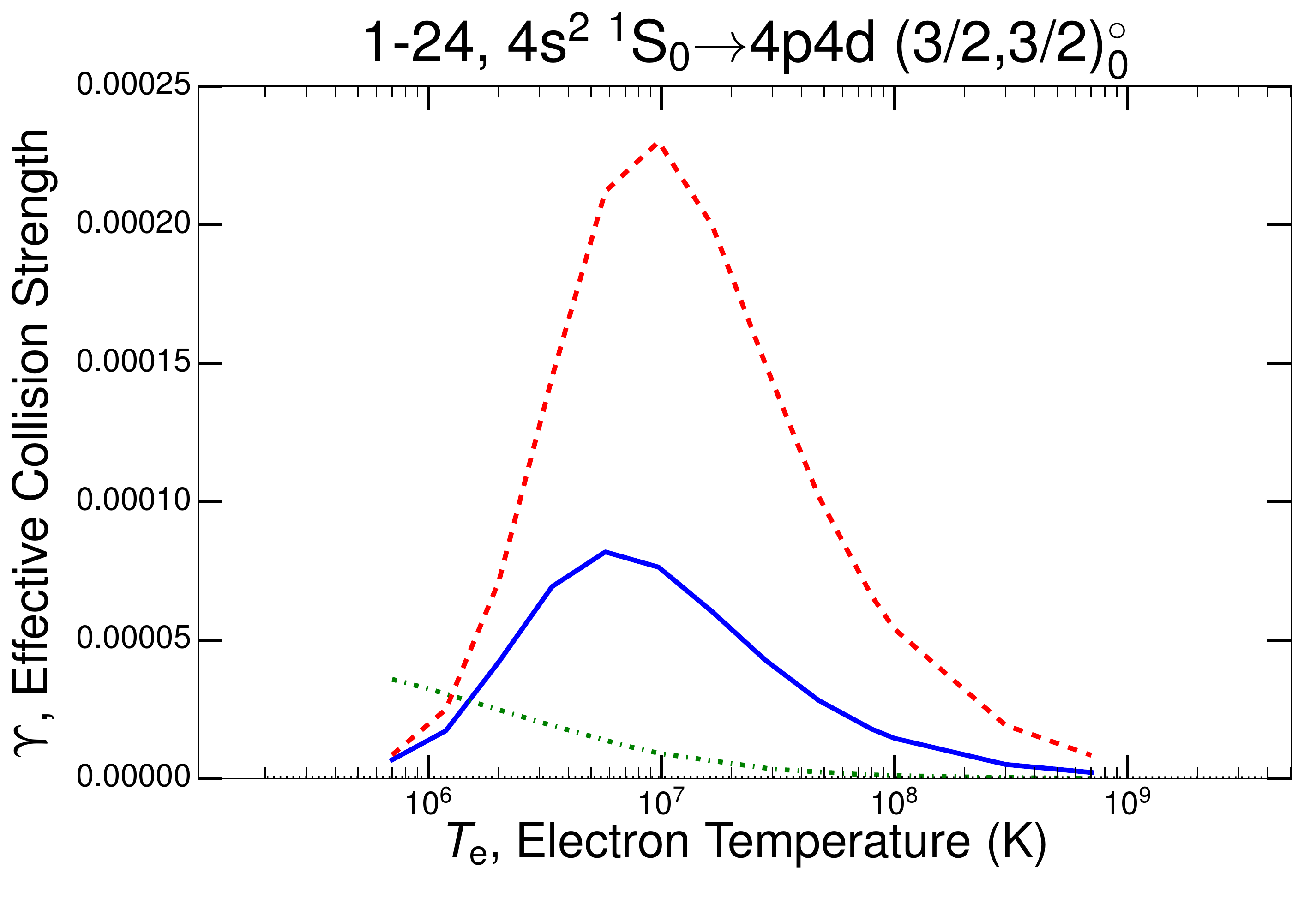}
            \end{center} \vspace{-0.75\baselineskip} \caption{}
            \label{fig:BGups1-24} \end{subfigure}
        \caption{ Collision strength, $\Omega$, and effective collision
            strength, $\Upsilon$, results for the three transitions presented
            by Ballance and Griffin in \cite{ballance2007}. Figures
            \subref{fig:BGom1-2}, \subref{fig:BGom1-3}, and
            \subref{fig:BGom1-24} display the convolution of the present
            $\Omega$ data with a 2.205 Ryd (30 eV) Gaussian function; this
            `smoothes' the dense resonance peaks while still retaining the
            information about where the peaks are strongest, making
            interpretation and viewing easier. The dashed (red) line is for the
            undamped data, and the solid (blue) line for the damped data.
            Figures \subref{fig:BGups1-2}, \subref{fig:BGups1-3}, and
            \subref{fig:BGups1-24} show the present $\Upsilon$ data
            (\helvet{DARC pres} and \helvet{DARC pres damp}) along with the present \AS\ DW
            (\helvet{AS DW}) results and the corresponding Ballance and
            Griffin (\helvet{BG}) results \cite{ballance2007}. Refer to
            the legend in \subref{fig:BGups1-2} for the line styles
        corresponding to each data set.} 
        \label{fig:BGcollision} \end{adjustwidth}
\end{figure}

\begin{figure}[htpb] \begin{adjustwidth}{-0.5in}{-0.5in}
        \begin{subfigure}{0.5\linewidth} \begin{center}
                \includegraphics[width=\linewidth]{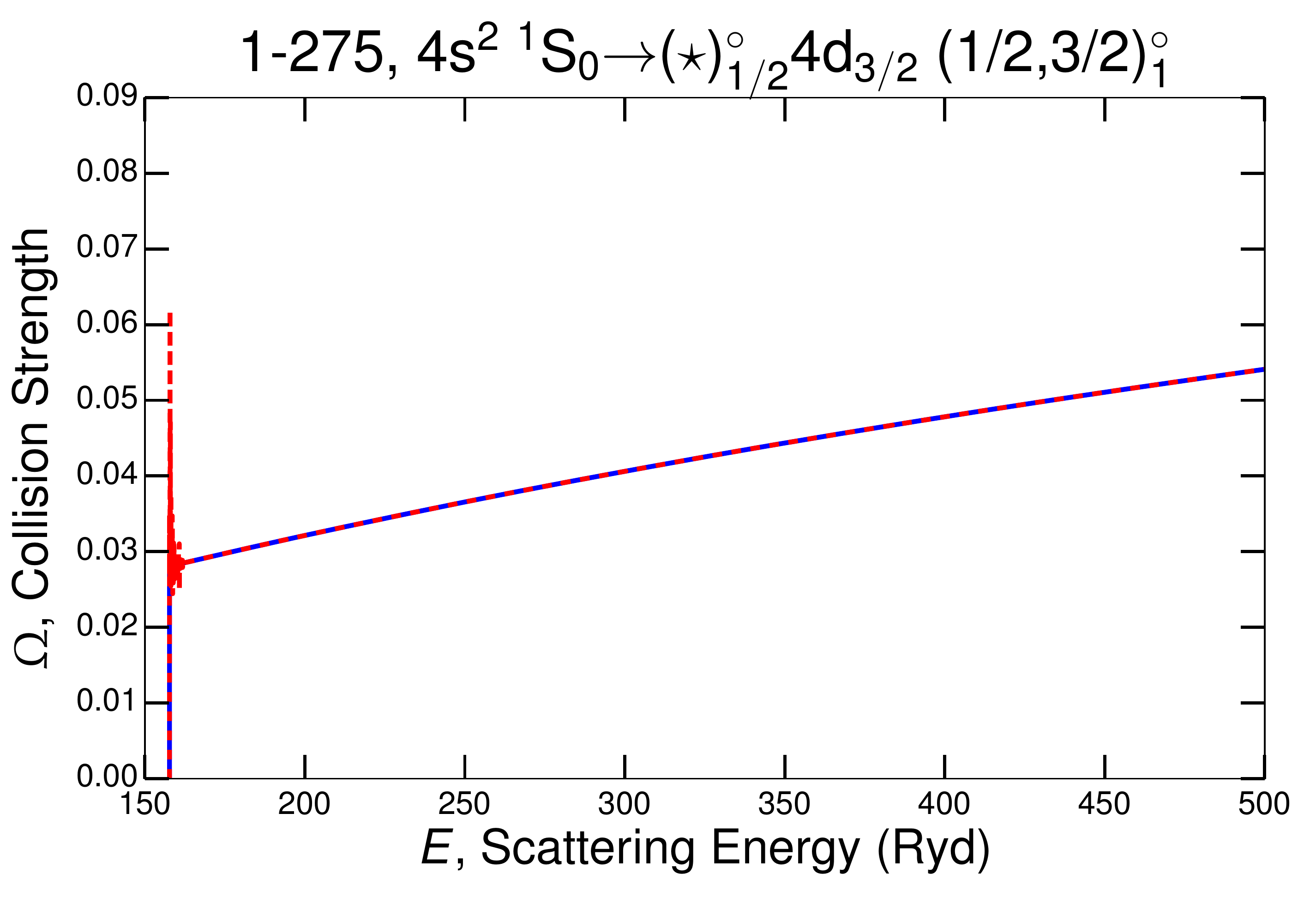}
            \end{center} \vspace{-0.75\baselineskip} \caption{}
            \label{fig:om1-275} \end{subfigure}
        \begin{subfigure}{0.5\linewidth} \begin{center}
                \includegraphics[width=\linewidth]{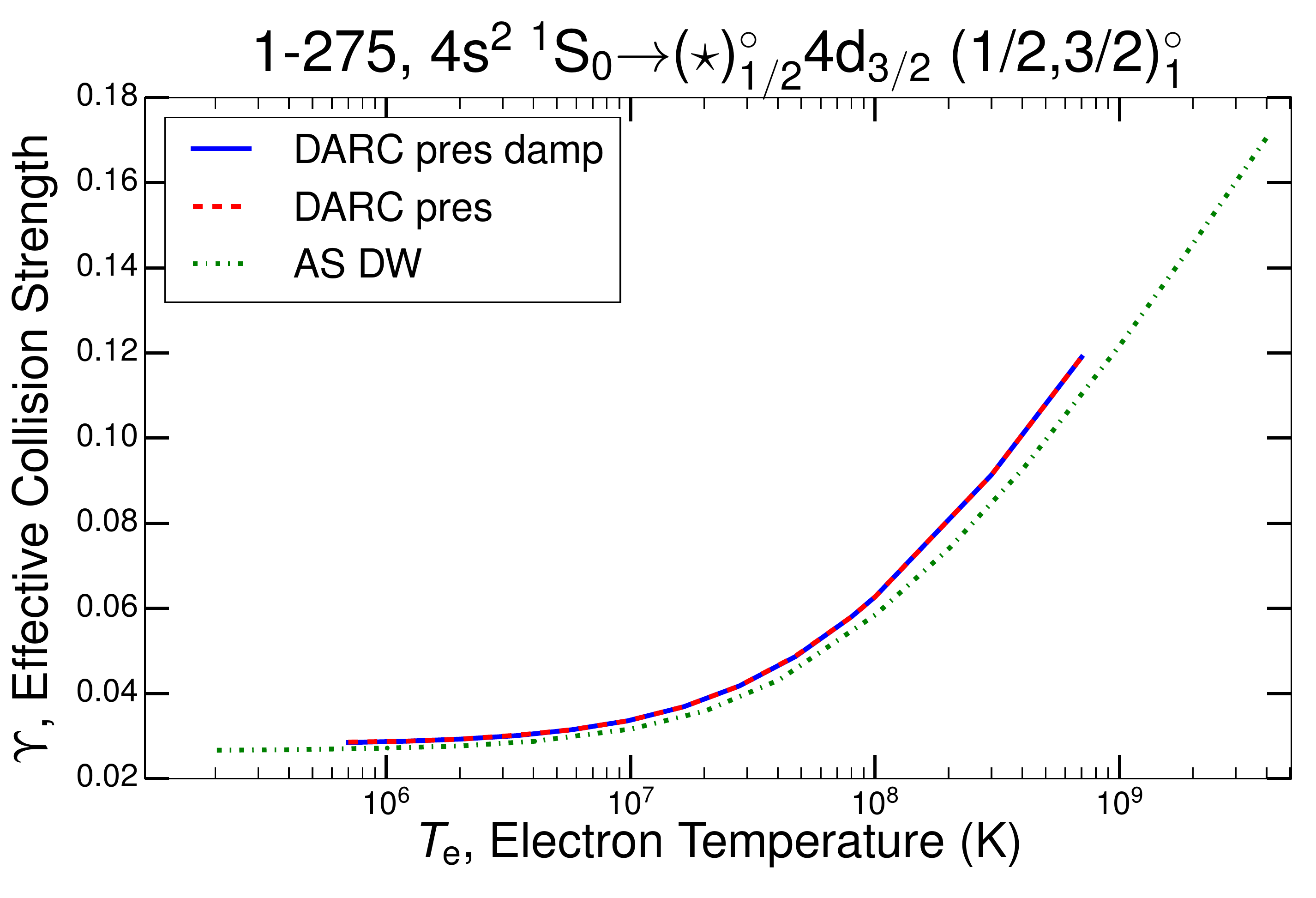}
            \end{center} \vspace{-0.75\baselineskip} \caption{}
            \label{fig:ups1-275} \end{subfigure} \\
        \begin{subfigure}{0.5\linewidth} \begin{center}
                \includegraphics[width=\linewidth]{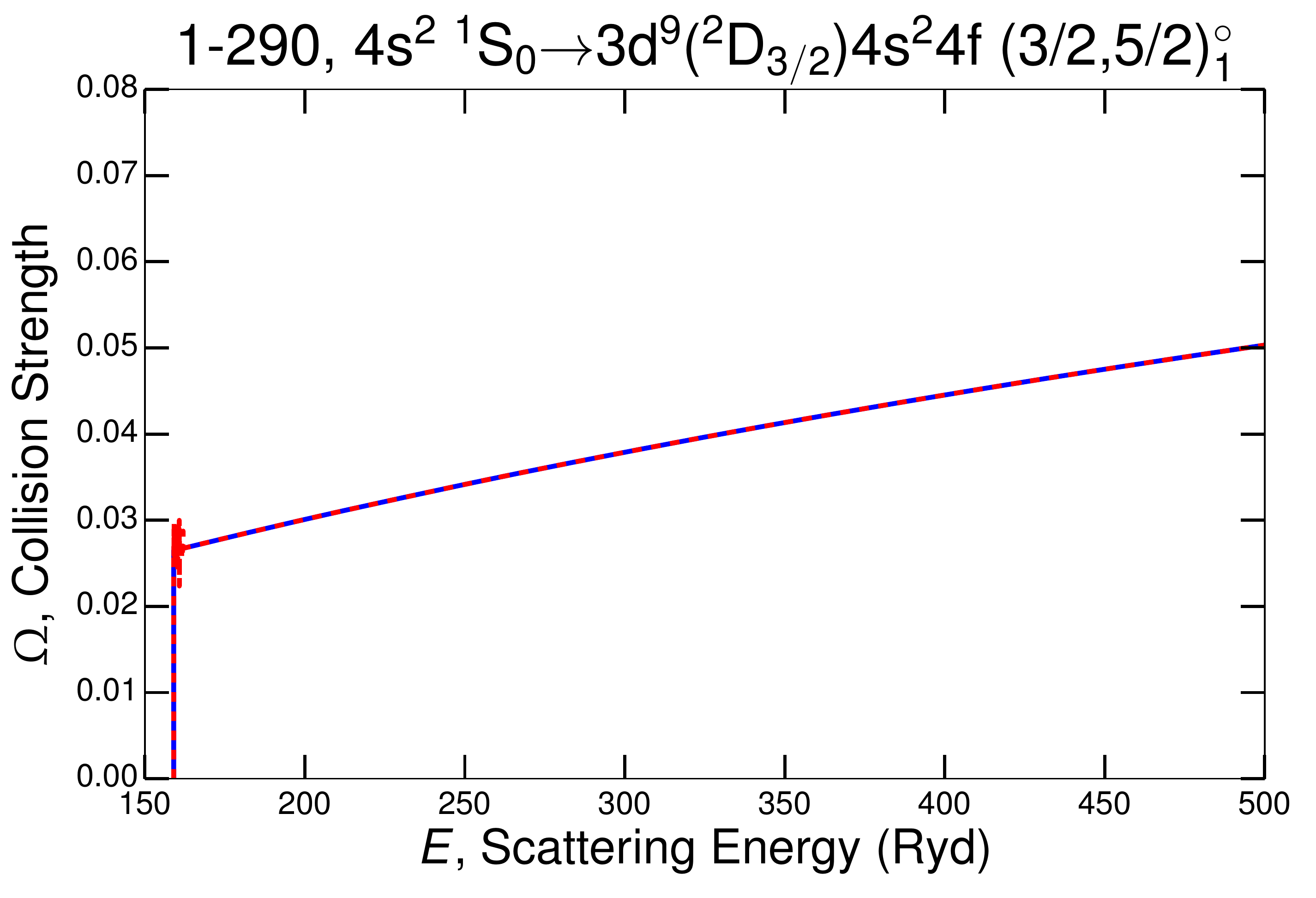}
            \end{center} \vspace{-0.75\baselineskip} \caption{}
            \label{fig:om1-290} \end{subfigure}
        \begin{subfigure}{0.5\linewidth} \begin{center}
                \includegraphics[width=\linewidth]{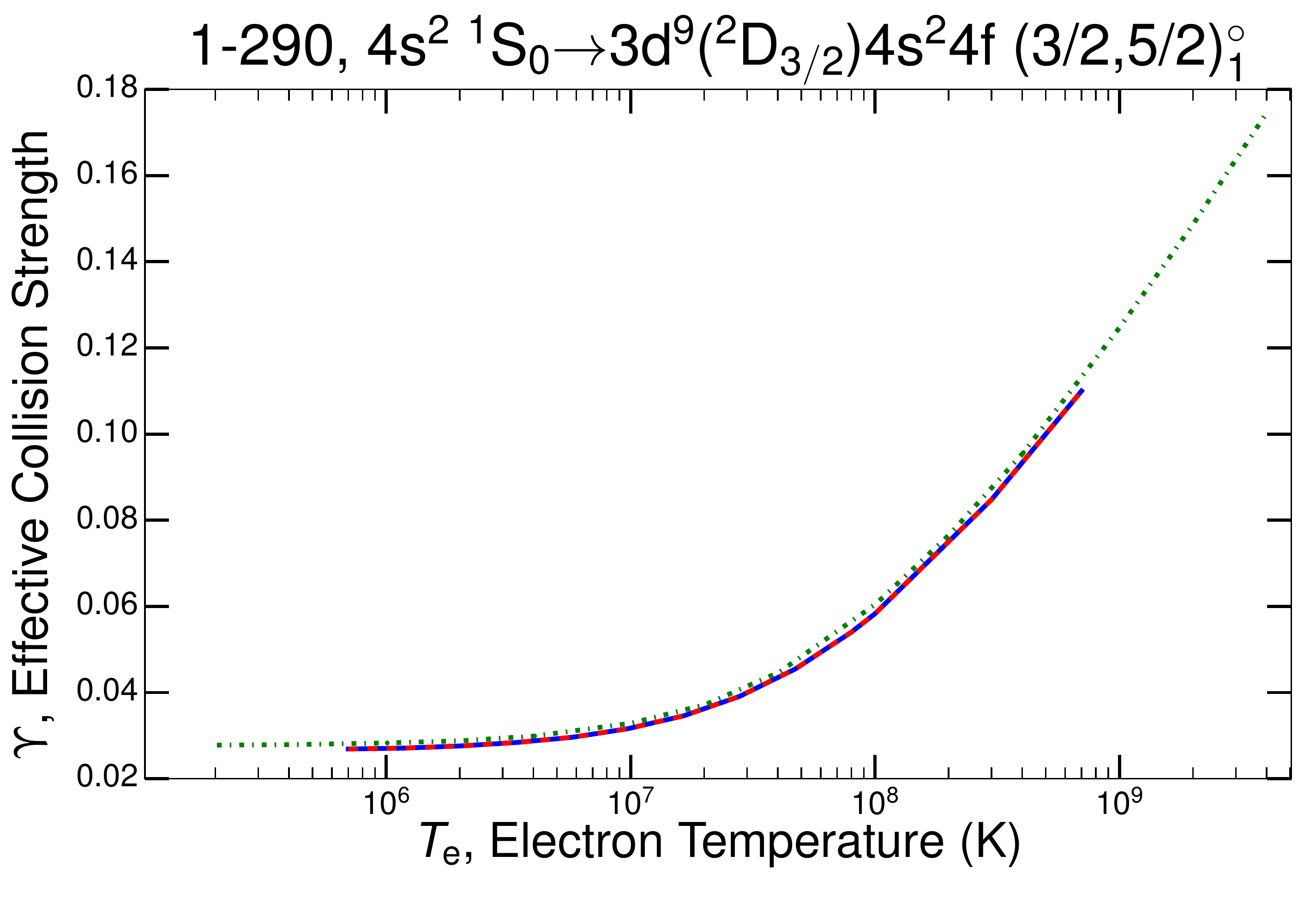}
            \end{center} \vspace{-0.75\baselineskip} \caption{}
            \label{fig:ups1-290} \end{subfigure} \\
        \begin{subfigure}{0.5\linewidth} \begin{center}
                \includegraphics[width=\linewidth]{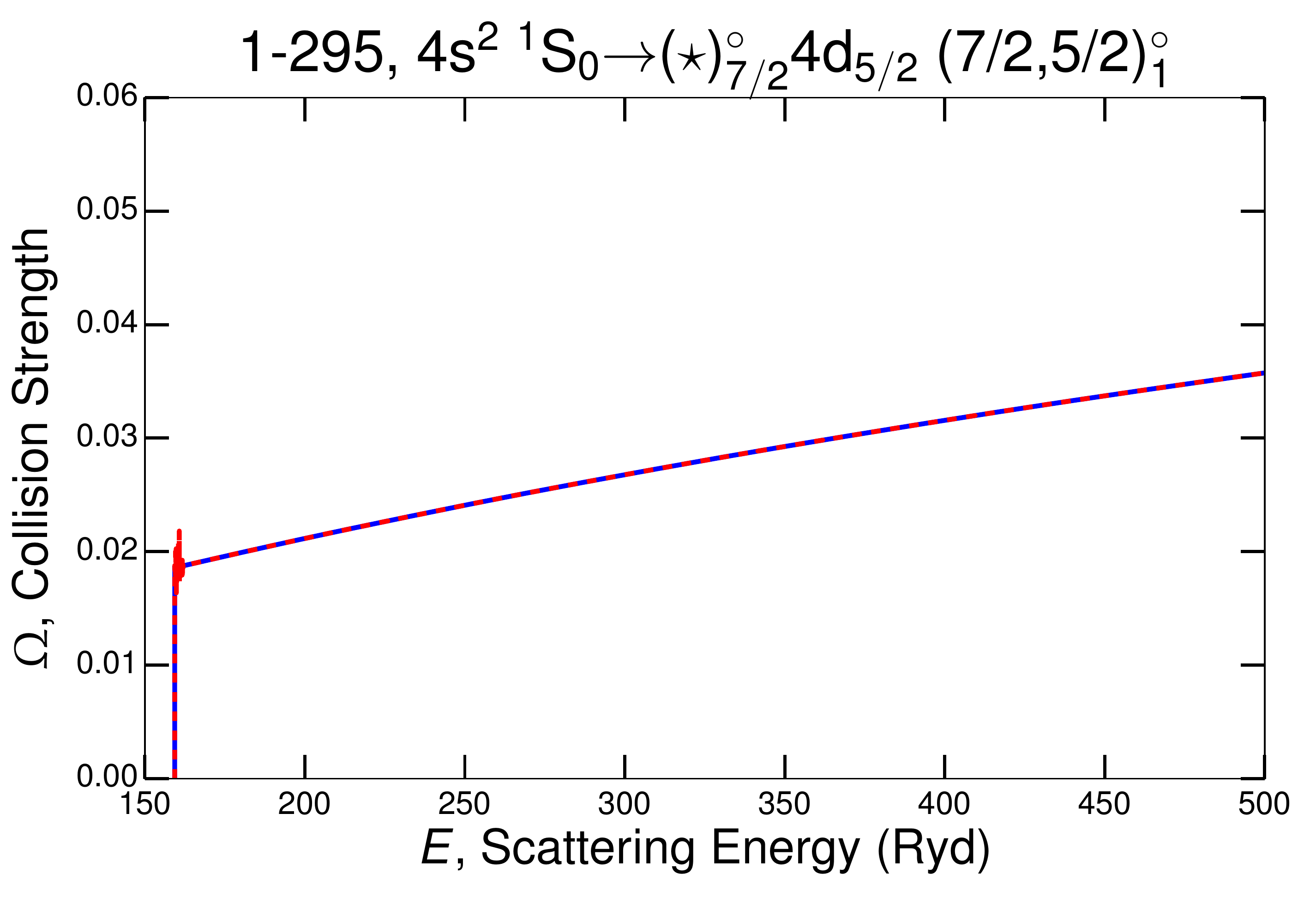}
            \end{center} \vspace{-0.75\baselineskip} \caption{}
            \label{fig:om1-295} \end{subfigure}
        \begin{subfigure}{0.5\linewidth} \begin{center}
                \includegraphics[width=\linewidth]{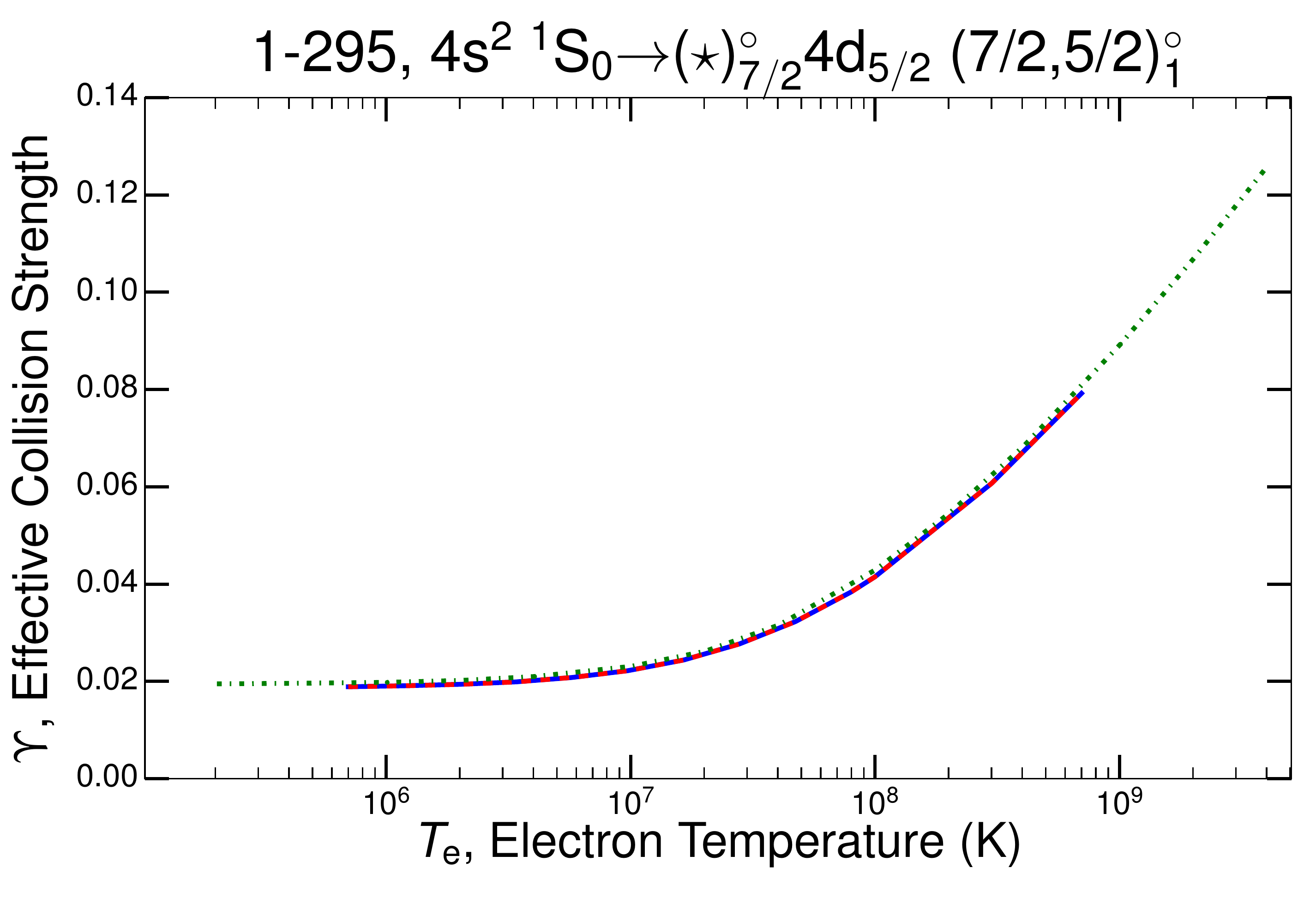}
            \end{center} \vspace{-0.75\baselineskip} \caption{}
            \label{fig:ups1-295} \end{subfigure}
        \caption{ Present results for the dominant 3d-subshell transitions in
            the transition arrays, \transarraya\ and \transarrayb. In contrast
            to figure \ref{fig:BGcollision}, \subref{fig:om1-275},
            \subref{fig:om1-290}, and \subref{fig:om1-295} are the `raw'
            $\Omega$ data sets that have not been convoluted; no convolution is
            required for these transitions because of the limited resonance
            structure. Again, the dashed (red) line is for the undamped data,
            and the solid (blue) line for the damped data. Figures
            \subref{fig:ups1-275}, \subref{fig:ups1-290}, and
            \subref{fig:ups1-295} display the $\Upsilon$ data for both the \darc{}
            and \AS\ DW calculations. Refer to the legend in
            \subref{fig:ups1-275} for the corresponding line styles. In the
        level specifications, the substitution, $\star \equiv $
        (3d$^9$($^2$D$_{5/2}$)4s$_{1/2}$)$^{\circ}_2$4p$_{3/2}$, is used.}
    \label{fig:important_collisions} \end{adjustwidth} \end{figure}

Moving now to the collision problem, a sample of the data from our \darc{} and
\AS\ DW calculations is provided in figures \ref{fig:BGcollision} and
\ref{fig:important_collisions}, and figure \ref{fig:BGcollision} also contains
data from the Ballance and Griffin calculations \cite{ballance2007} for
comparison \footnote{The energy levels, radiative rates, and effective
collision strengths from the present work are available in the adf04 file
format on the OPEN-ADAS website:
\url{http://open.adas.ac.uk/detail/adf04/znlike/znlike_mmb15][w44ic.dat}}.
This data is provided in the form of collision strengths and effective
collision strengths. The dimensionless collision strength, $\Omega(i,j)$, for
the transition between atomic states $i$ and $j$, is related to the
cross-section, $\sigma(i \rightarrow j)$, by
\begin{equation}
	\sigma(i \rightarrow j) = \frac{\pi a_0^2 I_{\mathrm{H}}}{g_i k_i^2} 
	\Omega(i,j) \,,
    \label{eq:omega}
\end{equation}
where $g_i$ is the statistical weight of the initial state, $k_i$ the
wavenumber of the incident electron, $a_0$ denotes the Bohr radius and
$I_{\mathrm{H}}$ is the ionization potential of the hydrogen atom in 
the units used for $k_i^2$.

The effective collision strength, $\Upsilon_{ij}$, is the thermal average of the
collision strength, typically a Maxwellian average that is used in the present
work:
\begin{equation}
   \Upsilon_{ij} = \int^{\infty}_{0} \Omega(i, j)  e^{( - \epsilon_j \slash
   kT_\mathrm{e})} \mathrm{d} ( \epsilon_j\slash kT_\mathrm{e}) 
    \label{eq:upsilon}
\end{equation}
where $\epsilon_j$ is the final energy of the scattering electron,
$T_\mathrm{e}$ the electron temperature, and $k$ denotes Boltzmann's constant.
The Maxwell-Boltzmann distribution is non-relativistic, and relativistic
effects become significant for $T_{\mathrm{e}} \gtrsim 20$ keV $\approx 2.3
\times 10^{8}$ K, relevant to the electron temperatures expected at ITER.
In keeping with ADAS convention, we do not apply any relativistic corrections to
the electron distribution functions used to produce the $\Upsilon_{ij}$ values
in this work. The relativistic Maxwell-J\"uttner distribution only requires the application of
a simple multiplicative factor to the Maxwell-Boltzmann $\Upsilon_{ij}$ values. 

Damping effects are apparent in both the collision strengths and effective
collision strengths in figure \ref{fig:BGcollision}, and our
\AS\ DW results are always less than the \darc{} results. This should be
expected since our DW does not include resonance contributions to the effective
collision strengths, which are certainly present for these
transitions.  However, the high energy behaviour of the DW results does
approach that of the \darc{} results as would be expected.

There are obvious differences of the damped effective collision strengths between the
present results and the Ballance and Griffin results for transitions 1--2 and
1--24,
figures \ref{fig:BGups1-2} and \ref{fig:BGups1-24} respectively. Both of these
transitions are non-dipole ($J:0\rightarrow 0$) and comparatively small in
magnitude; therefore, damping effects and any differences in the CC expansion
tend to be more pronounced. Our lack
of a full damping treatment could explain the discrepancies; however, one must
first compare the undamped data to resolve the true origin of any differences.
Unfortunately, the undamped Ballance and Griffin results are only presented in
graphical form in their paper and the original data files are not available
\cite{ballance2014}. Furthermore, only data for the
damped \emph{effective} collision strengths are available, not the damped
collision strengths. A visual comparison with the plots in the Ballance and
Griffin paper is still useful. Comparing our undamped effective collision
strengths with those of Ballance and Griffin, one still observes large
differences: our results are larger by about the same factor as in the damped
case. Any differences in the undamped \emph{effective} collision strengths must
be due to differences in the resonant structure of the undamped collision
strengths. Indeed, comparing our collision strengths in figures
\ref{fig:BGom1-2} and \ref{fig:BGom1-24} with the Ballance and Griffin
collision strengths, there are intensity peaks present in our results that are
not present in theirs, a direct indication that there are additional
intermediate resonances in our CC expansion. For example, transition 1-2 will
have the resonance 3d$^9$4s$^2$4p$nl$ available in our calculations but not in
Ballance and Griffin's.  Combining this and the observation that the relative
amount of damping in our results is comparable to the Ballance and Griffin
results --- inferred again from visual inspection --- it is reasonable to conclude
that the differences observed here are most likely due to the differences in
the CIs and CC expansions and not differences in the treatment of radiation
damping. Moreover, discrepancies due to varying resonant enhancement between
calculations should be less pronounced in strong dipole allowed transitions,
and this is exactly what is observed for the dipole 1-3 transition in figures
\ref{fig:BGom1-3} and \ref{fig:BGups1-3}.

\begin{figure}[htpb] \begin{adjustwidth}{-0.5in}{-0.5in}
        \begin{subfigure}{0.5\linewidth} \begin{center}
                \includegraphics[width=\linewidth]{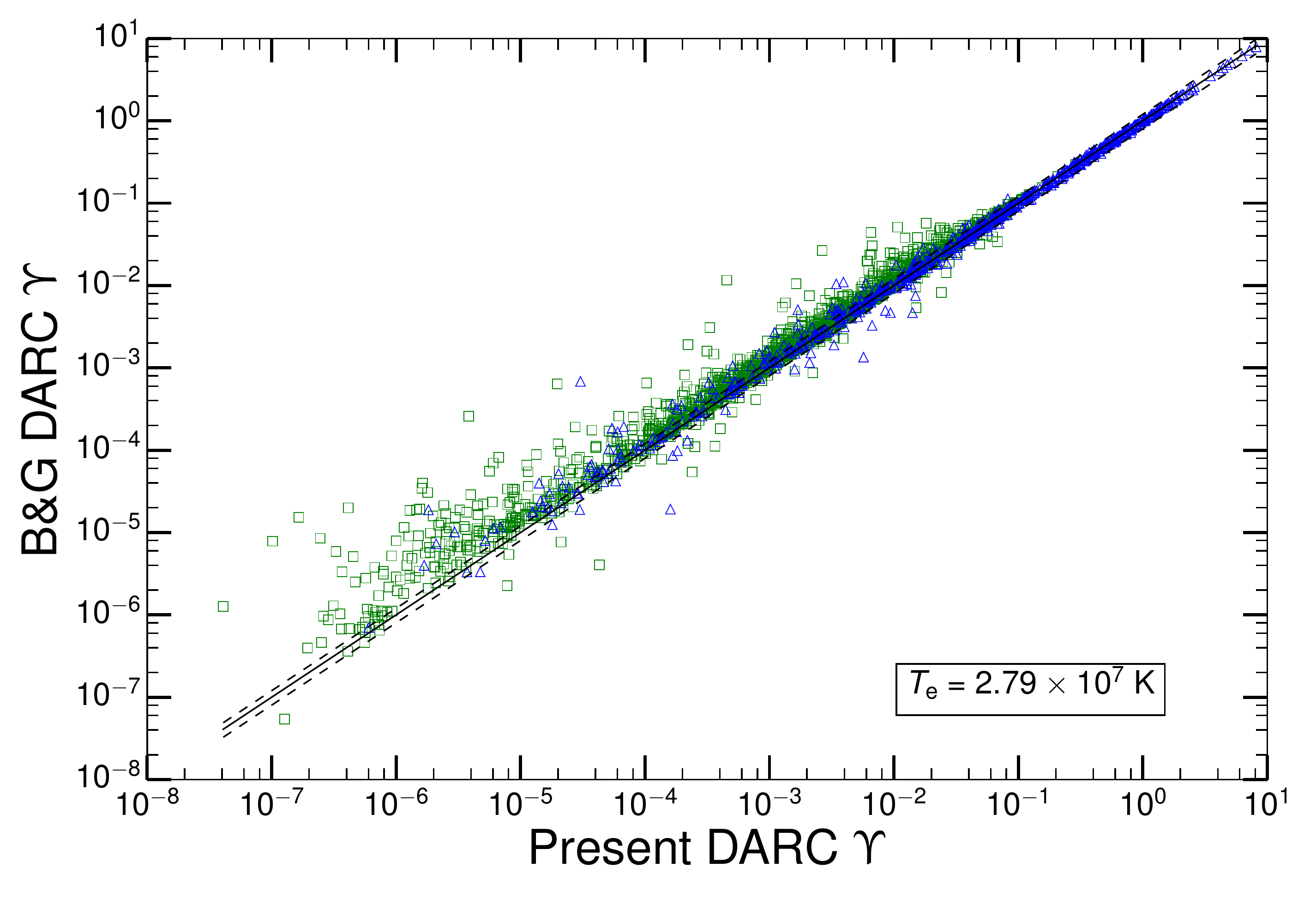}
            \end{center} \vspace{-0.75\baselineskip} \caption{}
            \label{fig:linplot1_all} \end{subfigure}
        \begin{subfigure}{0.5\linewidth} \begin{center}
                \includegraphics[width=\linewidth]{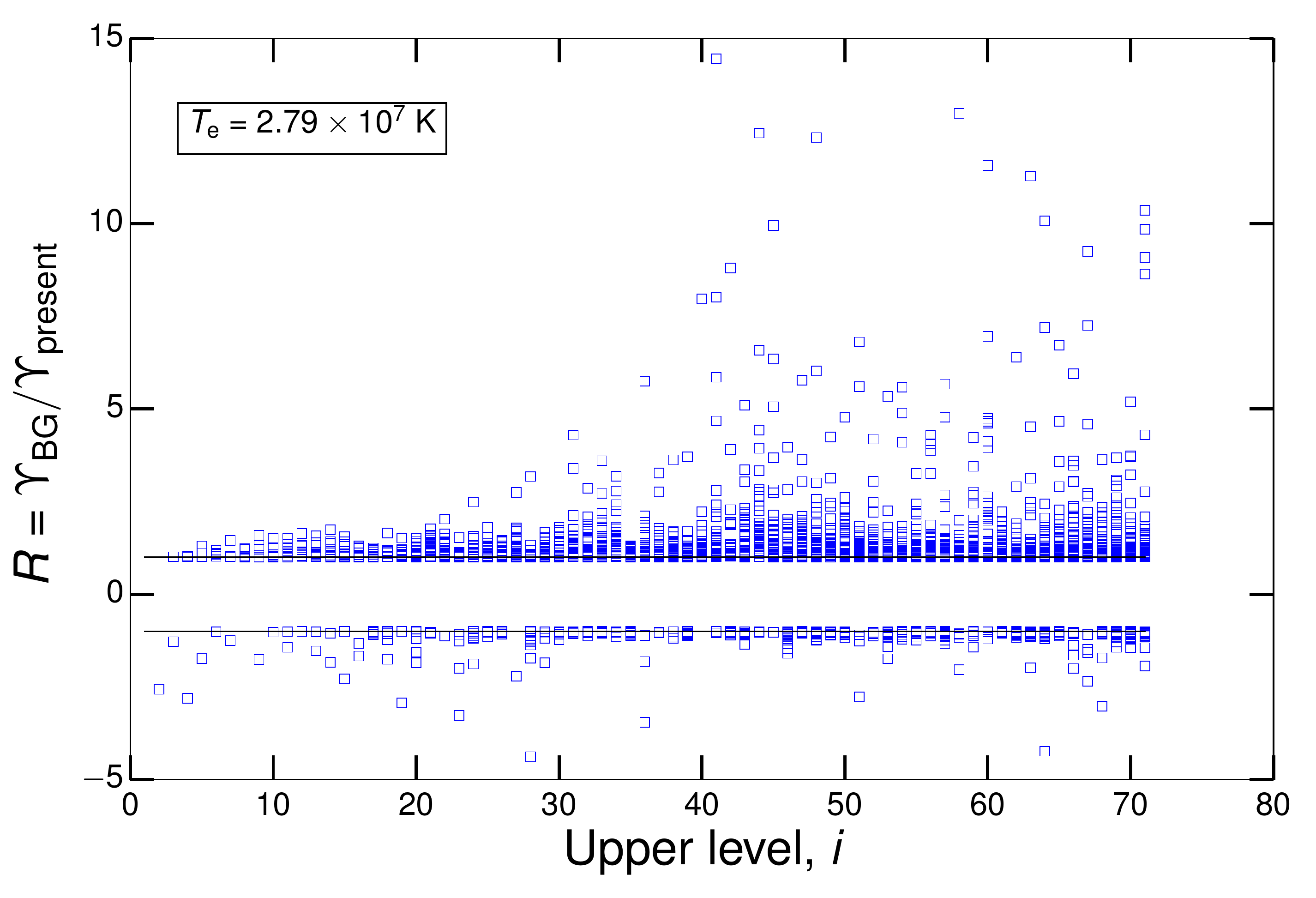}
            \end{center} \vspace{-0.75\baselineskip} \caption{}
            \label{fig:ratioplot1} \end{subfigure} \\
        \begin{subfigure}{0.5\linewidth} \begin{center}
                \includegraphics[width=\linewidth]{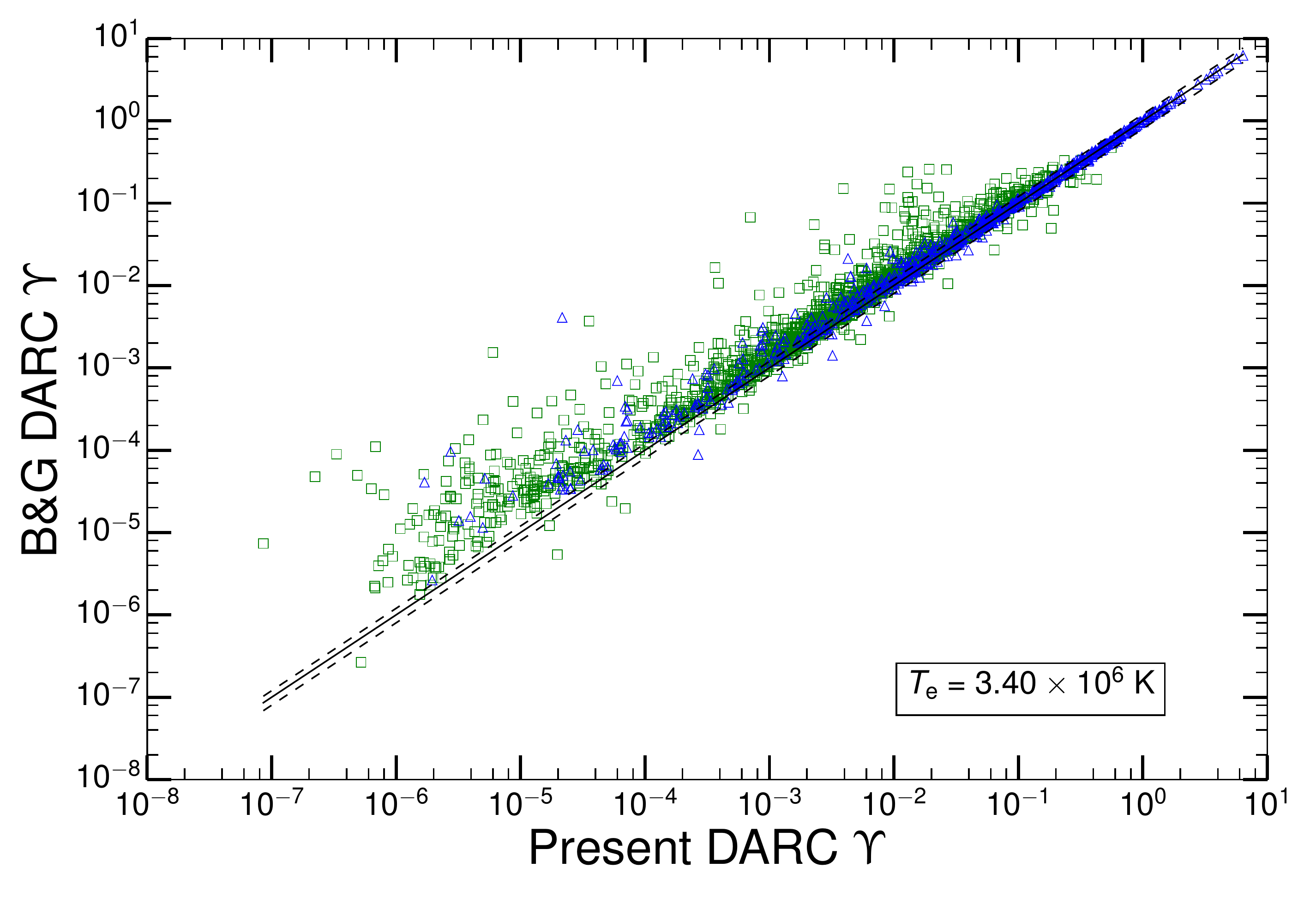}
            \end{center} \vspace{-0.75\baselineskip} \caption{}
            \label{fig:linplot2_all} \end{subfigure}
        \begin{subfigure}{0.5\linewidth} \begin{center}
                \includegraphics[width=\linewidth]{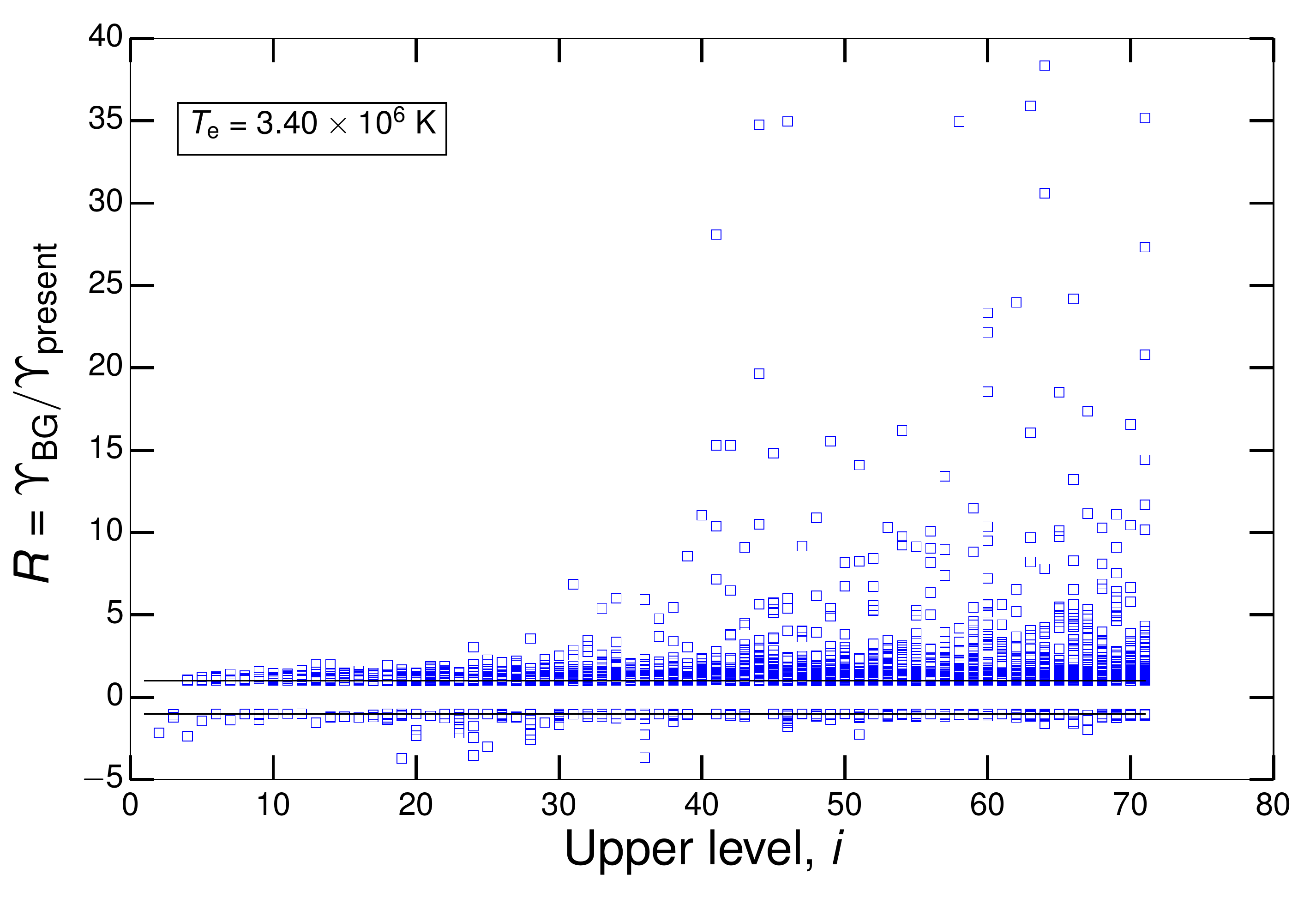}
            \end{center} \vspace{-0.75\baselineskip} \caption{}
            \label{fig:ratioplot2} \end{subfigure}
        \caption{Comparison --- \subref{fig:linplot1_all} and
            \subref{fig:linplot2_all} --- and ratio --- \subref{fig:ratioplot1} and
            \subref{fig:ratioplot2} --- scatter plots of effective collision
            strength values, $\Upsilon$, from the two primary calculations:
            Ballance and Griffin's (B\&G) fully damped \darc{} versus the present,
            partially damped \darc{}. The temperature at which the $\Upsilon$
            values are being sampled is indicated by the boxed value on each
            plot. For the comparison plots, \subref{fig:linplot1_all} and
            \subref{fig:linplot2_all}, the (blue) triangles denote dipole
            transitions, and the (green) squares denote non-dipole transitions.
            The dotted lines demarcate the 20\% error region around the $y=x$
            line, and the percentage of points within the error regions are as
            follows: \subref{fig:linplot1_all} all = 63\%, dipole = 82\%,
            non-dipole = 56\%; \subref{fig:linplot2_all} all = 44\%, dipole =
            68\%, non-dipole = 35\%. For the ratio plots,
            \subref{fig:ratioplot1} and \subref{fig:ratioplot2}, the binary
            positive or negative behaviour of the ratio is defined by $R =
            \Upsilon_{\mathrm{BG}} / \Upsilon_{\mathrm{present}}$ if
            $\Upsilon_{\mathrm{BG}} > \Upsilon_{\mathrm{present}}$ or $R = -
            \Upsilon_{\mathrm{present}} / \Upsilon_{\mathrm{BG}}$ if
            $\Upsilon_{\mathrm{BG}} < \Upsilon_{\mathrm{present}}$. The ratio
            is plotted versus the upper level, $i$, of the transition in each
            case.} 
            \label{fig:lin_plots}
        \end{adjustwidth} 
\end{figure}

Since these are only two cases, it is not possible to apply this conclusion in
general, and it would be impractical to analyze every transition in this
manner: there are 2843 intersecting transitions for the two calculations.
However, a slightly larger subset of about 15 transitions was analyzed in
similar detail, and the same conclusion was reached: our undamped effective
collision strengths tend to agree quite well with those of Ballance and Griffin
for strong transitions, but weaker transitions display variable levels of
agreement.  Still, this is not enough evidence to extrapolate our conclusion,
so a broader scope technique must be used.  Our approach was to select
temperatures of interest and then compare the effective collision strength
values from the two calculations for all intersecting transitions. Graphically,
this results in the comparison scatter plots presented in figures
\ref{fig:linplot1_all} and \ref{fig:linplot2_all}, one at a temperature near
that of peak abundance for \Wff ($\approx 3\times 10^7$ K) and the other at a
lower temperature. The intersecting levels involved in these transitions have
an index cut-off of $i = 71$, corresponding to the last 3d$^{10}$4d4f level.
Figure \ref{fig:config_energies} displays that above this configuration, the
energy level distributions do not intersect, and therefore there are no
overlapping transitions involving levels above this cut-off.

Our limited damping treatment compared to Ballance and Griffin means our
collision data should be systematically \emph{larger}, and this would manifest
as a statistically significant number of points lying below the $y=x$ line.
However, figures \ref{fig:linplot1_all} and \ref{fig:linplot2_all} display the
exact opposite: what appears to be a significant number of points above the
$y=x$ lines and so a systematic trend towards our $\Upsilon$ values having
comparatively \emph{smaller} magnitudes. Because the density of points in the
vicinity of the $y=x$ line is not readily estimated, it cannot be immediately
concluded that this is a statistically significant trend. Calculating the
fraction of points within an uncertainty region of 20\% around the $y=x$ line
can elucidate the situation, and the results of this calculation are presented
in the caption of figure \ref{fig:lin_plots}. The values of 63\% and 44\% for
the all transitions cases indicate that although there is reasonable agreement
between most points at these temperatures, a significant portion do lie outside
the uncertainty region. Additionally, plotting the ratio of the effective
collision strengths, $R = \Upsilon_{\mathrm{BG}} /
\Upsilon_{\mathrm{present}}$, versus a relevant independent variable as in
figures \ref{fig:ratioplot1} and \ref{fig:ratioplot2} can reveal important
systematic trends. Both of these plots show a clear asymmetry of higher
$\Upsilon$ values from the Ballance and Griffin calculations. Hence, the
significance of the systematic trend is supported.

Since the systematic trend is the opposite to what was expected, there must be
another, more significant systematic effect involved other than our limited
radiation damping treatment. From the observation of no systematic deviation in the
dipole line strengths in section \ref{sub:Structure Data}, it is deduced that
the systematic difference cannot be caused directly by differences in the atomic
structure. Several indicators suggest that this other
systematic effect must be additional resonant enhancement for low to intermediate
scattering energies in the Ballance and Griffin calculations. Firstly, the
comparison plots in figures \ref{fig:linplot1_all} and \ref{fig:linplot2_all}
both show that the trend towards larger $\Upsilon_{\mathrm{BG}}$ values is
relatively greater for weaker transitions.  The non-dipole transitions, because
they tend to be weaker, display a greater susceptibility to the trend,
supported by the lower error region percentages and a visibly larger spread of
values. Juxtaposing figures \ref{fig:linplot1_all} and \ref{fig:linplot2_all},
which only differ by the sampling temperature, reveals that the trend of larger
$\Upsilon_{\mathrm{BG}}$ values is enhanced at lower electron temperature, an
observation that is also true for figures \ref{fig:ratioplot1} and
\ref{fig:ratioplot2}. The preceding observations support the claim of
additional resonant enhancement because resonances tend to affect weaker,
non-dipole transitions to a larger degree and even more so at lower \Te. 

Secondly, it is seen from the ratio plots in figures \ref{fig:ratioplot1} and
\ref{fig:ratioplot2} that the $\Upsilon_{\mathrm{BG}}$ values are increasingly
large compared to ours as the index of the upper level, $i$, increases. The
upper level is relevant for resonant enhancement considerations because it
restricts the possible levels that can be involved in the intermediate ($N+1$)
resonant states. As the upper level of a transition approaches the level
intersection cut-off of $i = 71$ ($E \approx 8 \times 10^6$ cm$^{-1}$ in figure
\ref{fig:config_energies}), the transition will increasingly only have access
to resonances involving levels that are discrepant between the calculations.
Consequently, the tendency for $\Upsilon$ values to disagree more at higher $i$
that is observed in figures \ref{fig:ratioplot1} and \ref{fig:ratioplot2} is
consistent with the proposition of discordant resonant enhancement. 

However, this now begs the question why it is that the Ballance and Griffin
results have systematic, additional resonant enhancement, especially
when the present calculations include a larger number of levels. The answer must derive
from the differing structure of the CC expansions and thus the differing atomic
energy level distribution that is summarized in figure
\ref{fig:config_energies}. The non-intersecting, $n=5$ energy levels in the
Ballance and Griffin calculation are immediately above the dashed-line
threshold; hence, these levels will be more accessible for resonance formation
if the electron distribution functions peaks close to the excitation energy of
the transition under consideration. In contrast, the 3d-hole configurations
lie $\sim 60$~Ryd higher, as do resonances with the same $n$-value.
Furthermore, 3 of these 4 configurations have a strong dipole $4\mathrm{p}, 4\mathrm{f}
\rightarrow 3\mathrm{d}$ type-I radiation damping transition. Finally, some
common initial configurations --- 4p$^2$, 4p4f, 4d$^2$, and 4d4f --- have no single
electron promotions to our 3d-hole resonances, unlike Ballance and Griffin
where resonances can be formed by promotion to $n=5$. 

One point should be clear from the preceding discussion: it is the composition
of the CI and CC expansion that most influences the behaviour of the collision
data being compared. Indeed, it is still possible that our calculations neglect
a large amount of damping, which would be hidden by the cancellation of the two
systematic effects; however, this is unlikely given the analysis of figure
\ref{fig:BGcollision}. The objective of including consideration of the soft
x-ray, 3d-subshell transitions had necessarily shaped the CI/CC expansion used
in our calculations, and so differences with other calculations should be
expected. In the end, a true assessment of the merits of these two primary
calculations can only be obtained through the application of the data in the
atomic population modelling to follow.

Figure \ref{fig:important_collisions} shows the collision data for the strongest 
three 3d-subshell transitions. Because of the strength of these $E1$ transitions,
resonances appear to be unimportant and the behaviour due to direct Coulomb excitation
dominates. Such observations are supported by a sharp jump in the collision
strengths at the energy threshold of each transition. The limited number of
resonance peaks is due to the fact that the upper levels in these transitions
are close to the highest energy level included in our calculation, meaning
there are comparatively few intermediate resonant states available.
Furthermore, good agreement is observed between the \AS\ DW results
and the \darc{} effective collision strengths. Again, this can be accounted for by
the relative sparsity and small magnitude of resonances for these transitions.
One might be tempted to conclude that it would be simpler and
less time consuming to have only used the DW results; however, it is difficult
to predict whether the results will still be similar following atomic
population modelling.  So it is prudent to carry all available results --- present
\darc{}, \AS\ DW, Cowan PWB, and Ballance and Griffin \darc{} --- forward and
assess any differences following the final analysis. 

\subsection{Atomic Population Modelling} 
\label{sub:Modelling_Results}

As noted in Section \ref{sec:Introduction}, determination of the total radiated
power loss from \Wff\ is one of the desirable outputs from atomic population
modelling. The excitation line power coefficient for a transition,
$j\rightarrow k$, is defined by
\begin{equation}
    P_{L,1, j \rightarrow k} = \Delta E_{jk}
    \PEC^{(\mathrm{exc})}_{1, j\rightarrow k}\ ,
    \label{eq:plt}
\end{equation}
which has units of (W cm$^3$) and is simply the relevant \PEC\ multiplied by
the energy difference between the levels involved, $\Delta E_{jk}$. The total
excitation line power coefficient, $P_{LT,1}$, is the sum of the $P_{L,1, j
\rightarrow k}$ over all possible transitions and is directly proportional to
the total radiated power loss of the ionization stage. Although the
\PEC{}s and power coefficients give much of the same information,
\PEC{}s are preferred in spectroscopic applications while power coefficients
are needed for estimates of radiated power loss. Both are employed in the
subsequent analysis and are largely interchangeable in cases where general
conclusions about a transition are being sought.

\begin{figure}[htpb]
    \centering
    \begin{subfigure}{0.45\linewidth}
        \begin{center}
            \includegraphics[width=\linewidth]{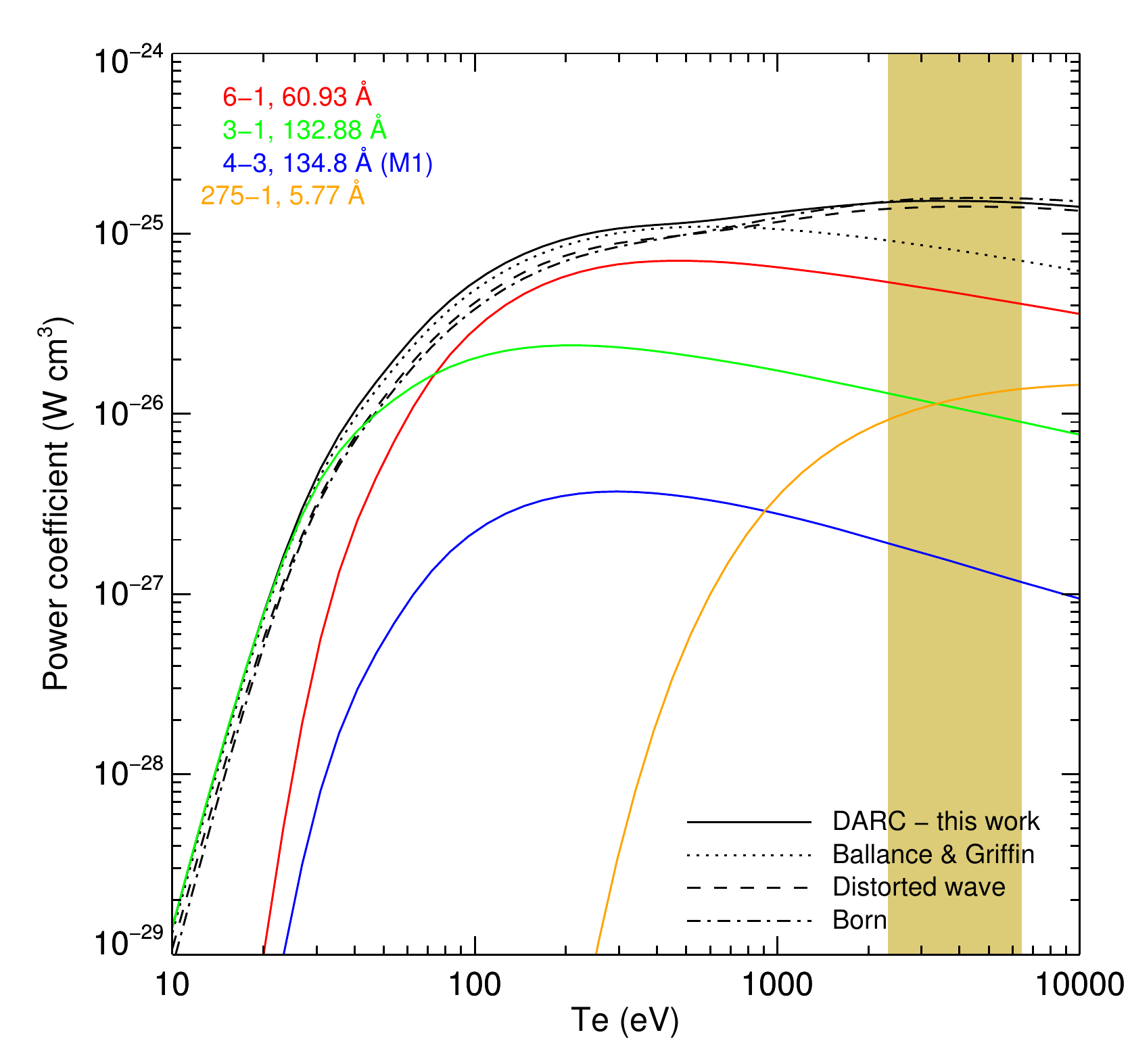}
        \end{center}
        \vspace{-0.75\baselineskip}
        \caption{}
        \label{fig:power_coeffs}
    \end{subfigure}
    \begin{subfigure}{0.45\linewidth}
        \begin{center}
            \includegraphics[width=\linewidth]{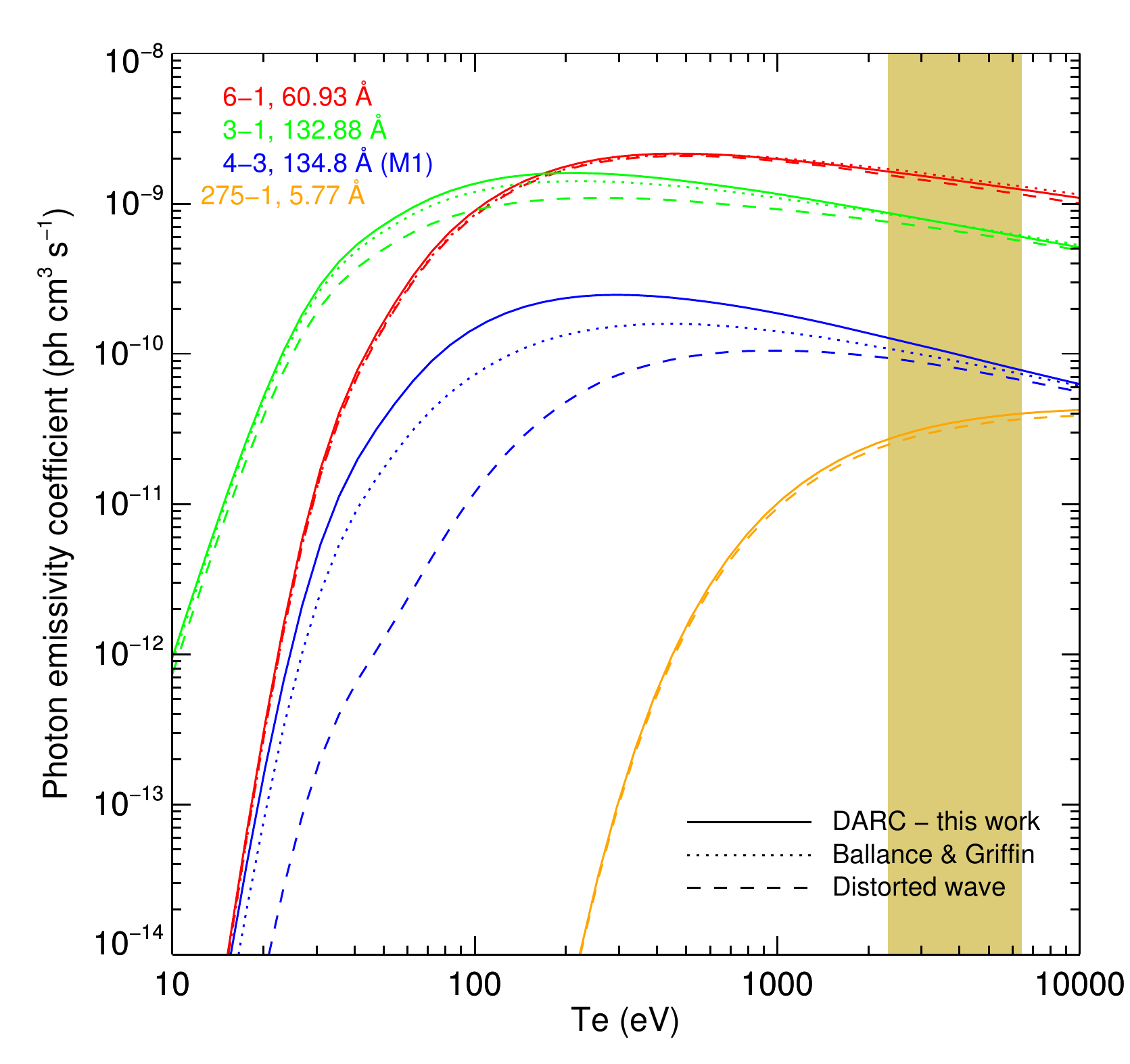}
        \end{center}
        \vspace{-0.75\baselineskip}
        \caption{}
        \label{fig:PECvsT}
    \end{subfigure}
    \caption{\PLT{}, \PL{}, and \PEC{} values derived from the relevant
    fundamental datasets for \Wff{} versus electron temperature, \Te{}. The
    shaded vertical bar represents the $T_{\mathrm{e}}$ range where the
    fractional abundance of \Wff{} in the coronal equilibrium approximation is
    greater than 0.1. \subref{fig:power_coeffs} shows the total excitation line
    power coefficients, \PLT{}, as the enveloping (black) lines, and these
    have been calculated for the four $\Upsilon$ datasets with line styles
    indicated in the figure: the Ballance and Griffin \darc{} and the present \darc{}, 
    \AS\ DW, and Cowan PWB. A sample of the strongest and most
    relevant contributing individual lines from the present \darc{} work have
    been emphasized (coloured) and labelled.  \subref{fig:PECvsT} displays the
    \PEC{} lines for the corresponding \PL{} lines in \subref{fig:power_coeffs}.
    The line styles denote different datasets as labelled in the
    figure: Ballance and Griffin's \darc{} and the present \darc{}
    and \AS\ DW.  \emph{Note}: there are no Ballance and Griffin results
    for the  275--1 (5.77\AA) \PEC{} line. The indices from our \grasp{}
    calculation are used --- see tables \ref{tab:E_levels} \ref{tab:rad_data}.}
    \label{fig:rad_coefficients}
\end{figure}

The total excitation line power coefficients from the various calculations are
plotted versus electron temperature in figure \ref{fig:power_coeffs}, along
with a selection of relevant, contributing $P_{L,1, j \rightarrow k}$ from
our present \darc{} work. Observing the individual \PL{} values, the dominant
transition across most of the $T_{\mathrm{e}}$ range is unsurprisingly the
dipole allowed 6--1 (60.93 \AA); however, towards lower \Te\ the VUV 3--1
(132.88 \AA) transition is stronger due to its lower energy difference.  Most
importantly for this work, the strongest line from the open 3d-subshell
transition arrays is the highlighted 275--1 (5.77 \AA) transition.
It is the value of the power coefficient at peak abundance temperatures that is
of most concern, and a critical observation is that the 275--1 3d-subshell line
contributes an equal amount to the total radiated power as does the VUV 3--1
line in this region. 

The salient feature of the \PLT{} lines in figure \ref{fig:power_coeffs} is the
departure of the Ballance and Griffin result from the other calculations at
high \Te, commencing just before the demarcated region of peak abundance. What
causes this behaviour is evident from the individual \PL{} lines, just
discussed: the 275--1 (5.77 \AA) line, which is not included in the Ballance
and Griffin calculations, rises to a 50:50 power contribution with the strong
VUV 3--1 (132.88 \AA) transition in the peak abundance region. Omission of this
line along with others of comparable magnitude in the \transarraya\ and
\transarrayb\ transition arrays leads to the relative reduction in the \PLT{}
seen in the Ballance and Griffin results. Otherwise, the \PLT{} values from the
other calculations, both of which include at least some of the important
3d-hole configurations, agree well across the given \Te\ domain with no
relative errors over 50\% and convergence at high \Te{}, notably in the shaded
region of peak abundance. This reiterates a common theme: the primacy of the
configurations included in the collision calculation and subsequent modelling.
Without appropriate consideration of the 3d-subshell transitions, a large
contribution to the radiated power from \Wff{} will be missed, reaffirming our
decision to focus attention on these transitions. 

Figure \ref{fig:PECvsT} provides a more detailed point of comparison between
the calculations by showcasing the \PEC{}s for the same transitions as the
individual \PL{} lines in figure \ref{fig:power_coeffs}. Although the \PEC{}s
and \PL{} only differ by an energy factor, it is interesting to note the effect
that this has upon the importance of the 275--1 (5.77 \AA{}) line; the \PL{}
values are comparatively higher because of the large energy difference between
level 275 and 1. Agreement between the theories in figure \ref{fig:PECvsT} is
quite good for the strong dipole allowed transitions (3--1, 6--1, 275--1), and
the moderate discrepancy between the \darc{} and DW results for the 3--1 line
can be explained through application of the zero density limit expression in
\eref{eq:lowdens3}. This provides a good approximation in the present
circumstance because density effects on level populations are largely absent
until $N_{\textrm{e}}\approx 10^{16}$ cm$^{-3}$.  The dominant $A_{i\rightarrow
j}$ value in the sum of \eref{eq:lowdens3} is $A_{3\rightarrow 1}$ by many
orders of magnitude, and so the $A_{3\rightarrow 1}$ in the numerator will be
effectively cancelled. Thus, it must be variation in the excitation rate
coefficient, $q_{1\rightarrow 3}^{\mathrm{e}}$, that causes differences in the
\PEC{} values --- recall, excitation from the ground dominates in the zero
density limit. Indeed, the \AS{} DW $\Upsilon_{1\rightarrow 3}$ values are
systematically lower than the corresponding \darc{} values because of the
absence of resonant enhancement; this explains why the DW \PEC{} is also lower
across the temperature range.

On the other hand, the spin-changing, $M1$, 4--3 transition displays notable
differences between all of the calculations, but the \PEC{} values do
eventually converge at high \Te{}. Again, these differences can be understood
through the use of the zero density limit for the \PEC{}, and just as above, the
contributions from the radiative transition probabilities cancel due to the
dominance of the $A_{4\rightarrow 3}$ value. The $\Upsilon_{1\rightarrow
4}$ values for the various calculations reproduce the ordering of
the 4--3 \PEC{} lines in figure \ref{fig:PECvsT}: the \AS{} DW $\Upsilon_{1\rightarrow
4}$ are less than both of the \darc{} results because of the
absence of resonances, and our \darc{} $\Upsilon_{1\rightarrow
4}$ are larger than Ballance and Griffin's for less obvious reasons. The trend
of relatively larger Ballance and Griffin $\Upsilon$ values observed in section
\ref{sub:Collision Data} in no way means that our $\Upsilon$ values for a
particular transition cannot be larger as is the case here; however, the cause
of this is indeterminable without the ability to look at the Ballance and Griffin
$\Omega$ data. 

There are several conclusions relevant to radiated power loss from the observations of 
figure \ref{fig:rad_coefficients}. First, the importance of the soft x-ray 3d-subshell 
transitions: the \PLT{} lines from figure
\ref{fig:power_coeffs} clearly show that neglecting the \transarraya{} and
\transarrayb{} transition arrays will greatly reduce predictions of radiated
power loss from \Wff{}. Thus, these transition arrays must be included
in the collision calculations upon which any effort to model radiated power
loss is built. Second, there is substantial evidence that the omission of
transitions involving
the 3d$^{10}$4$l$5$l^{\prime}$ configurations (henceforth, $n=5$ transitions) has little 
effect upon the \PLT{} values. The Cowan PWB result, which does include some $n=5$
transitions, does not deviate significantly from the present \darc{} nor the
\AS{} DW result. Furthermore, Ballance and Griffin collision data for the
$n=5$ transitions was merged into our present
\darc{} data, and a negligible effect upon the modelled quantities in figure
\ref{fig:rad_coefficients} was observed. The \PEC{}s still agreed to within a few
percent except for the 4--3, $M1$ transition which agreed within 10\%. Even
though
this merging is not a replacement for a full calculation with all of the
relevant configurations, it strongly indicates that
the $n=5$ transitions are not essential for radiated
power loss considerations in general and therefore also for the
3d-subshell transitions. As discussed in section \ref{sub:Collision Data}, the 
\nfive{} configurations do provide additional resonant enhancement for lower
level transitions, and the effect of this in the context of population
modelling will require further investigation outside the current scope of the
present study.

Thirdly, the overall proximity between the present \darc{}, Cowan PWB, and
\AS\ DW results in figure \ref{fig:power_coeffs} propounds the
suitability of the non-close coupling theories as baseline descriptions of the
radiated power from \Wff{}. However, the precedent statement in no way
recommends that the more intensive \darc{} calculations are unnecessary. From a
detailed spectroscopic perspective, one must assess the suitability of a
particular dataset on a transition-by-transition basis, and the small number of
transitions presented in figure \ref{fig:rad_coefficients} do not allow any
generalizations to be made. Another technique is required.

Because \Wff{} is a heavy and relatively complex species, there are so many
transitions that describing it with individual line emissivities is
overwhelming and not useful. In response, we produce envelope lines, defined by
a vector of feature photon-emissivity coefficients (\FPEC{}), that
are composite features of many \PEC{} lines over a wavelength region. Suppose
the spectral interval of interest, $[\lambda_0, \lambda_1]$, is partitioned by
$N_p$ elements of the set, $\left\{ \lambda_i \equiv \lambda_0 +
i(\lambda_1-\lambda_0)/N_p : i = 0, \ldots, N_p - 1\right\}$, then the envelope
feature photon emissivity coefficient vector is defined as
\begin{equation}
    \FPEC{}_{1,i}^{(\textrm{exc})} = \sum_{j,k; \lambda_{j\rightarrow k} \in
[\lambda_0, \lambda_1]} \PEC{}_{1,j\rightarrow k}^{(\textrm{exc})}
\int_{\lambda_i}^{\lambda_{i+1}} \varphi_{j\rightarrow k}(\lambda)\dd \lambda
    \label{eq:FPEC}
\end{equation}
where $\varphi_{j\rightarrow k}(\lambda)$ is the normalized emission profile of
the spectrum line $j\rightarrow k$ that defines the line broadening.

\begin{figure}[htpb]
    \centering
    \includegraphics[width=\linewidth]{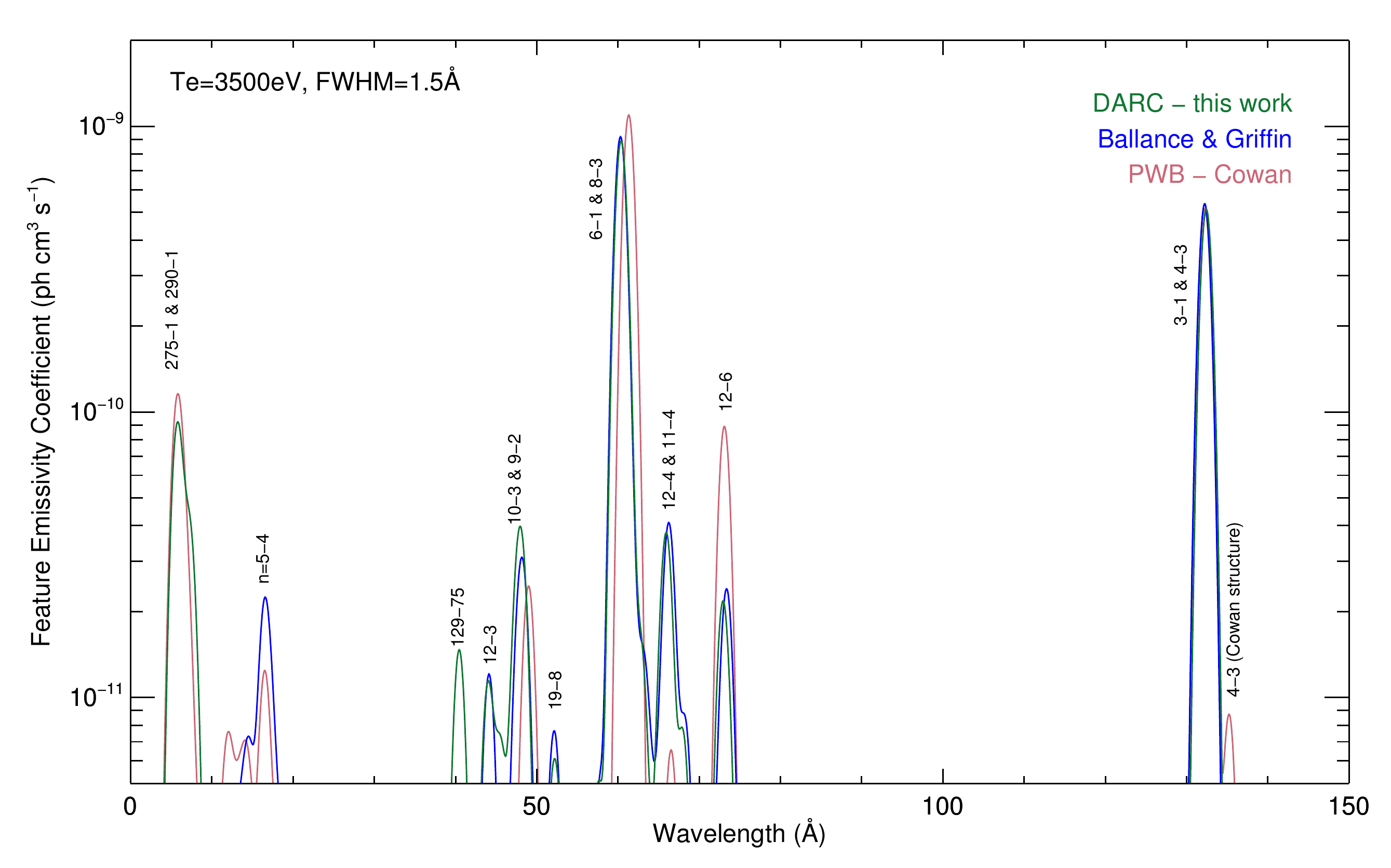}
    \caption{The envelope feature photon-emissivity coefficient, \FPEC{},
        vectors for various \Wff{} calculations plotted versus wavelength at
        $T_{\mathrm{e}} = T_{\mathrm{i}} = 3.5$ keV, where $T_{\mathrm{i}}$ is
        the ion temperature.  The calculations shown are those indicated in the
        top right, colour-coded legend: Ballance and Griffin's \darc{}, and the
        present \darc{} and PWB based on Cowan's code. The Doppler broadening
        by the velocity distribution of the radiating ions has been applied
        using the default Maxwellian distribution with $T_{\mathrm{i}} =
        T_{\mathrm{e}}$. In addition, the results were convolved with an ideal
        spectrometer instrument function with a FWHM of 1.5 \AA{}. The vertical
        labelling of the peaks denotes the transition(s) for the dominant
        excitation \PEC{}(s) within the feature; the indices from our
        \grasp{} calculation are used --- see tables \ref{tab:E_levels}
        \ref{tab:rad_data}.
    }
    \label{fig:FPEC}
\end{figure}

The spectral features resulting from the \FPEC{} vectors of the various \Wff{}
datasets are plotted in figure \ref{fig:FPEC}; portions of soft x-ray and VUV
regions are represented. As might be expected, the intensities of the features
which envelop strong transition lines agree well --- the peaks labelled by 6--1
\& 8--3 ($\sim 61$ \AA{}) and 3--1 \& 4--3 ($\sim 132$ \AA{}). However, the
6--1 feature does display some wavelength discrepancy. The Cowan PWB result
overestimates slightly compared to the two \darc{} results. For features of
less intense lines, the disagreements are larger: the Cowan PWB result differs
from the two \darc{} results by nearly an order of magnitude for both the 12--4
\& 11--4 ($\sim 66$ \AA{}) and 12--6 ($\sim 73$ \AA{}) features. Additionally,
the 10--3 \& 9--2 ($\sim 48$ \AA{}) peak exhibits both intensity and wavelength
discrepancies between all the calculations. Overall, figure \ref{fig:FPEC} also
clarifies the wavlength coverage of these three datasets. Of most relevance for
this work is that there is no Ballance and Griffin result for the 275--1 \&
290--1 ($\sim 7$ \AA{}) feature, which is the third most intense. Again, this
corresponds to the dominant soft x-ray, 3d-subshell transitions that we have
been concerned with throughout, and our \darc{} result is in close agreement
with the Cowan PWB. In addition, our \darc{} work has no data between 10 \AA{}
and 20 \AA{} corresponding to where the $n=5\textrm{--}4$ lines lie. 

The unifying message from the observations of figure \ref{fig:FPEC} is that
there are enough differences between the CC and non-CC calculations such that
applications in detailed spectroscopy could produce disparate results --- for
example, when calculating the line emissivity, $\varepsilon_{i\rightarrow k}$,
from \eref{eq:emissivity}. However, the two \darc{} results do agree very well
for overlapping spectral intervals. This further supports the conclusion above
that our neglect of the $n=5$ transitions has not significantly affected the
modelled results. A possible criticism
of this conclusion is that only strong emission lines are being considered in
figure \ref{fig:FPEC} and that differences between the datasets might become
more apparent for weaker lines. But this point is moot: the very fact
that these lines are weak and not part of this spectrum means they will not be
observable and so are irrelevant from an experimental standpoint. Therefore,
for both spectroscopic and radiated power applications, we recommend our
\darc{} adf04 file with the merged $n=5$ transition data from Ballance and Griffin. 


\section{Conclusion} 
\label{sec:Conclusion}

Fully relativistic, partially radiation damped, Dirac \rmatrix\ calculations
for the EIE of \Wff\ have been carried-out using the \grasp/\darc{} suite. The
energy levels, radiative rates, and effective collision strengths from the
present work are available in the adf04 file format on the OPEN-ADAS
website:
\url{http://open.adas.ac.uk/detail/adf04/znlike/znlike_mmb15][w44ic.dat}.
The primary objective and motivation for these calculations was to incorporate
both of the spectroscopically important transition arrays, \transarraya{} and
\transarrayb{}, which, to the best of our knowledge, had not been done until
now. Ultimately, any evaluation of our calculations must be made while keeping
this objective in mind. In addition, our \AS{} BPDW and Cowan PWB calculations
were conducted concurrently to provide baseline comparisons.

The inclusion of the configurations associated with the 3d-subshell transitions
required compromises to be made in the CI/CC expansion; configurations
3d$^{10}$4$lnl^{\prime}$ for $n>4$ were excluded due to computational limits.
Conversely, the Ballance and Griffin Dirac \rmatrix{} calculations with which
we compare included configurations for $n=5$ but did not open the 3d-subshell
to accommodate the 3d-subshell transitions. This difference in the CI/CC expansions
leads to a systematic difference between the $\Upsilon$ datasets which is likely
caused by an increase in resonant enhancement of the Ballance and Griffin results,
rather than being due to target structure or radiation damping variation.

Inevitably, evaluation of the differences in fundamental collision data is
performed through its application in atomic population modelling. From the
perspective of radiated power loss, it is clear from the \PLT{} and \PL{} lines
that the effect of the 3d-subshell transitions is far greater than any effects
due to the neglect of the $n=5$ transitions. Moreover, the non-CC calculations
provide a suitable baseline for radiated power loss estimates.
Spectroscopically, differences in the \FPEC{} spectra demonstrate that the
\rmatrix{} (CC) calculations are necessary for detailed applications, but the close
agreement of our \darc{} results with those of Ballance and Griffin further
supports the conclusion that omitting the $n=5$ transitions does not have a
large effect upon the modelled results. Indeed, it is the inclusion of the
3d-subshell transitions, which create a relatively strong spectral feature,
that is of greater import. In the future, it would be advantageous to extend
the present calculations to include the 3d$^{10}$4$l$5$l^{\prime}$ configurations so
as to unequivocally resolve the effect of the additional resonant enhancement
upon the lower lying transitions in the context of atomic population modelling.


\ack
MMB gratefully acknowledges the Natural Sciences and
Engineering Research Council (NSERC) of Canada and the Scottish University
Physics Alliance (SUPA) for providing his PhD funding.

\bibliography{references,myreferences}

\end{document}